\documentclass[12pt,tightenlines,nofootinbib]{revtex4-2}
\usepackage{amsfonts,amsmath,amssymb}
\usepackage{txfonts}
\usepackage{graphicx,color}
\usepackage{wrapfig,lipsum}
\usepackage{epstopdf}
\usepackage{hyperref}
\usepackage[shortlabels]{enumitem}
\usepackage{enumitem}
\usepackage{cancel}
\usepackage{mathrsfs,mathtools}
\newcommand{\secref}[1]{\autoref{1}.\nameref{1}}

\newcommand{\HP}{H \mkern -7.4mu P}

\newcommand\Alpha{\mathrm{A}}

\begin{document}

\title{Spaces and sequences in the hippocampus: a homological perspective}
\author{A. Babichev, V. Vashin$^1$, Y. Dabaghian$^1$} 
\affiliation{$^1$Department of Neurology, The University of Texas McGovern Medical School, 
	6431 Fannin St, Houston, TX 77030\\
$^{*}$e-mail: Yuri.A.Dabaghian@uth.tmc.edu}
\vspace{17 mm}
\date{\today}

\begin{abstract}
	Topological techniques have become a popular tool for studying information flows in neural networks. In 
	particular, simplicial homology theory is used to analyze how cognitive representations of space emerge
	from large conglomerates of independent neuronal contributions. Meanwhile, a growing number of studies
	suggest that many cognitive functions are sustained by serial patterns of activity. Here, we investigate
	stashes of such patterns using \textit{path homology theory}---an impartial, universal approach that does
	not require \textit{a priori} assumptions about the sequences' nature, functionality, underlying mechanisms,
	or other contexts.

	We focus on the hippocampus---a key enabler of learning and memory in mammalian brains---and quantify 
	the	ordinal arrangement of its activity similarly to how its topology has previously been studied in terms
	of simplicial homologies. The results reveal that the vast majority of sequences produced during spatial
	navigation are structurally equivalent to one another. Only a few classes of distinct sequences form an
	ordinal schema of serial activity that remains stable as the pool of sequences consolidates. Importantly,
	the structure of both maps is upheld by combinations of short sequences, suggesting that brief activity
	motifs dominate physiological computations.

	This ordinal organization emerges and stabilizes on timescales characteristic of spatial learning, 
	displaying similar dynamics. Yet, the ordinal maps generally do not reflect topological affinities---spatial
	and sequential analyses address qualitatively different aspects of spike flows, representing two 
	complementary formats of information processing. 
\end{abstract}
\maketitle
 
\newpage

\section{Introduction}
\label{sec:int}

One of the central tasks in neuroscience is to explain learning and memory through neuronal activity \cite{Koch}.
A key role in enabling these phenomena is played by the hippocampus---one of the brain's fundamental regions, 
both functionally and phylogenetically. Broadly, there are currently two perspectives on the organization of
hippocampal neuronal activity. On the one hand, it is believed that memory episodes may be encoded and invoked
by the firings of individual cells and their combinations, known as \textit{cell assemblies}\footnote{Throughout
the text, semantic highlights and terminological definitions are given in \textit{italics}. Key mathematical
terms are defined in Sec.~\ref{sec:met}}.

Specifically, hippocampal neurons, called place cells, learn to activate in specific areas within an 
environment---their respective place fields. Subsequent firings of these cells may be triggered not only by
the animal's physical visits to these fields but also by the recall of previous visits or anticipation of 
upcoming ones \cite{Crr,Ch2,Drg1,Drg2}. Suppressing the firing of select place cells using electrophysiological
or optogenetic tools can prevent the animal from entering corresponding fields, whereas their stimulation can 
prompt such visits \cite{Liu1,Gridchyn}. Curiously, hippocampal cells may also learn to selectively respond to
odors, visual cues, objects, memory episodes, and more, thus mapping out not only spatial but also mnemonic 
information \cite{Knudsen,Kolib,Wirth,Cao,Kutter}.

On the other hand, the hippocampus also enables animals to understand event sequences, plan series of actions,
learn cue sequences, and more \cite{Davachi,DuBrow,Fortin1}. For instance, rats with hippocampal lesions can 
recognize odors but struggle to memorize their order \cite{Fortin1}; lesioned monkeys have difficulty learning
new sequences of stimuli \cite{Murray,Fortin1}; and humans with hippocampal damage lose the ability to rank the
positions of objects in learned sequences \cite{Hsieh,Niel,Friston,Henin}. Physiologically, sequential learning
and recall are believed to be enabled by a series of consecutively firing cells and their assemblies, and hence
the temporal order of activity receives significant attention \cite{Kyle,Tzl,Syntax,Nicl}.

The mechanisms that produce either ordinal or spatial cognitive maps---and the principles that allow the 
hippocampus to sustain both functions---remain unknown. Generally, it can be argued that hippocampal 
representations are coarse, providing a rough-and-ready framework---with great computational speed, flexibility
and stability---into which geometric details, supplied by other brain regions, could be situated over time 
\cite{Poucet0}.
For example, memorized sequences can scale to accommodate different physical sizes and temporal durations,
adapting to specific tasks or environments \cite{Terada,Maurer,Mau}. Moreover, this scaling can be modulated
by cognitive and behavioral circumstances; for instance, items may be subjectively judged as ``closer" or 
``farther" apart, depending on the context \cite{Ezzyat}. 
Similarly, hippocampal cells preserve their relative order of spiking amidst gradual geometric transformations
of the navigated environment \cite{Gthrd,Ltgb,Wills,Trzky,eLife,Place,Blmnd}, which also points at a 
qualitative, topological nature of the encoded map \cite{eLife,PLoS,Rueck}. 

Over the last decade, several topological approaches have been used to model hippocampal maps and their
dynamics, based on simplicial \cite{Curto}, persistent \cite{Ghrist,Arai,Bso,Hof,CAs,MWind2,Eff,Rev,Kang},
and zigzag \cite{PLoZ,Replays} homology theories, along with other techniques \cite{Smith,CohnMap}. However,
in all these cases, constructing cognitive maps were viewed as assembling contributions from individual
spiking units, rather than building frameworks of sequential activity. To address the latter, we utilize an
alternative technique---path homology theory \cite{Grig1,Grig2,Grig3,Grig4,Grig5}, which qualitatively 
broadens the topological perspective on hippocampal activity. 

The paper is organized as follows. Section II discusses current topological models of hippocampal activity
and outlines path homology theory along with its connections to physiology. The results of path-homological
analyses of simulated neuronal dynamics are presented in Section III and discussed in Section IV.

\section{Topological methods}	
\subsection{Simplicial schema of cognitive map}

\begin{wrapfigure}{c}{0.75\textwidth}
	\centering
	\includegraphics[scale=0.75]{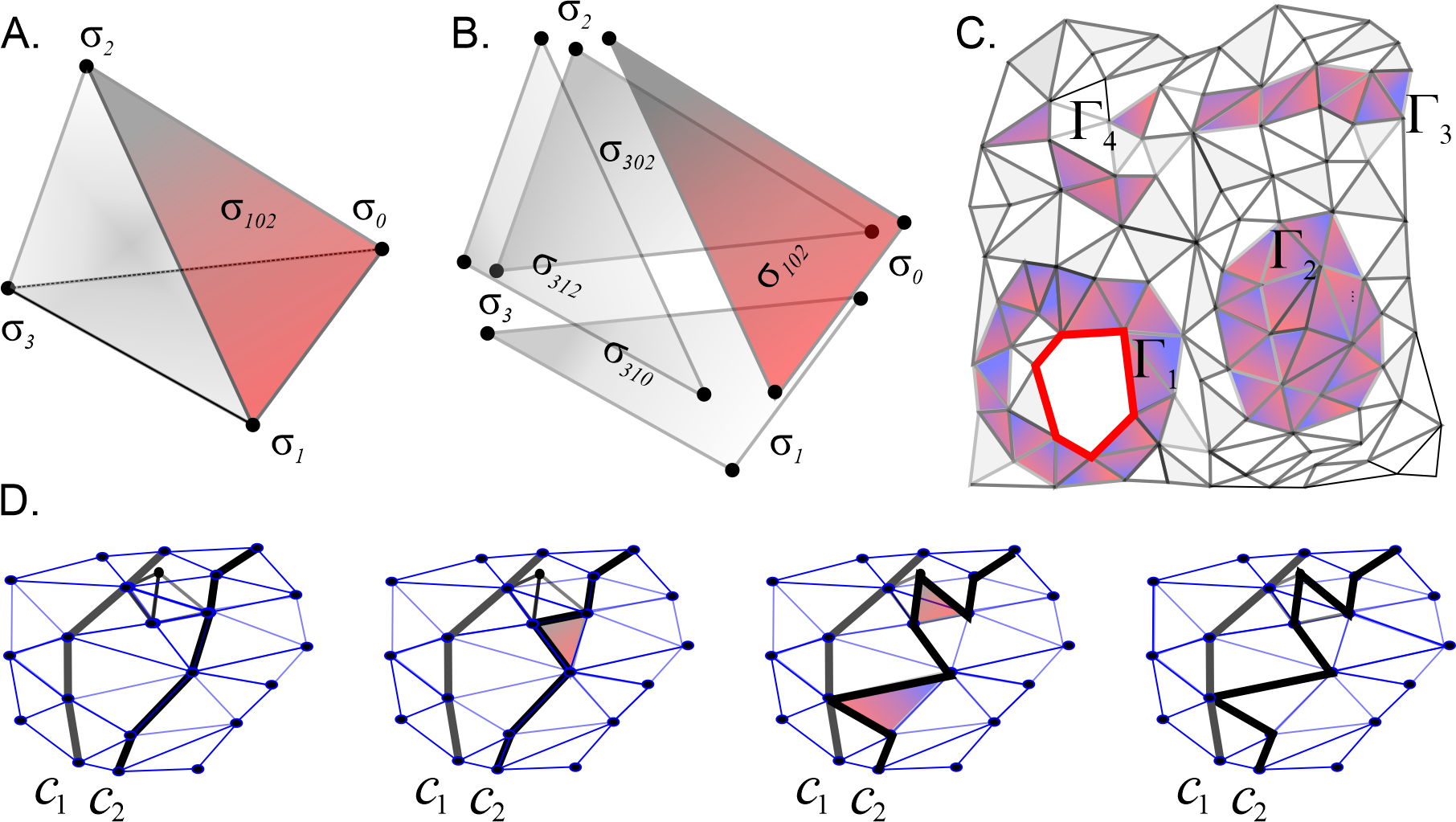}
	\caption{\footnotesize{\textbf{Simplexes and simplicial chains}.
			\textbf{A}. Simplexes are $k$-dimensional polytopes spanned by $k+1$ vertices. Shown is a 
			three-dimensional ($3D$) simplex---a tetrahedron.
			\textbf{B}. The boundary of a $3D$ simplex is a combination of its two-dimensional ($2D$) facets.
			\textbf{C}. Simplexes with matching boundaries form a simplicial complex. A chain in such a
			complex is an arbitrary combination of simplexes, counted with coefficients from a field 
			$\mathbb{K}$. Here colored simplexes illustrate $2D$ chains connected ($\Gamma_1$, $\Gamma_2$
			$\Gamma_3$) and disconnected ($\Gamma_4$), contractible ($\Gamma_2$, $\Gamma_3$, $\Gamma_4$)
			and	topologically nontrivial---a $1D$ cycle ($c_1$, red).
			\textbf{D}. Two homologous one-dimensional ($1D$) chains, $c_1$ and $c_2$, highlighted by
			thick black lines. The panels show a series of discrete deformations that transform one chain
			into the other, produced by ``snapping" segments of $c_2$ over the boundaries of adjacent
			(red) simplexes.
	}}
	\label{f1:simp}
	\end{wrapfigure}

A cell assembly is a transient group of neurons that work together\footnote{``...much like jazz musicians"
\cite{BuzBook}.} to elicit responses from downstream ``readout" neurons \cite{Syntax}. A given neuron,
$c_i$, may participate in multiple assemblies, meaning the cell assemblies are interconnected. Formally, an
assembly can be viewed as an abstract simplex (Fig.\ref{f1:simp}A), 
\begin{equation}
	\sigma_i \equiv [c_{i_0}, c_{i_1}, \ldots, c_{i_k}] \equiv \sigma_{i_0 i_1 \ldots i_k},
	\label{simplex}
\end{equation}
with the subsimplexes corresponding to independently activating ``subassemblies" of $\sigma_{i}$ 
\cite{Alexandrov,EdelHarer,ZmBook}. The combinations of cells activated by time $t$ form a simplicial 
coactivity complex, 
\begin{equation}
	\mathcal{T}(t) = \cup_{t_i < t} \sigma_i, 
	\label{tcomplex}
\end{equation}
which provides a collective representation of the information encoded by individual spiking units \cite{Rev}.
The shape of the coactivity complex reflects the overall, large-scale structure of information and exhibits
properties manifesting at the cognitive level. For instance, such complexes capture the shape of the navigated
 environment: temporally filtered persistent simplicial homologies of $\mathcal{T}(t)$ evolve to match the
 homological structure of the environment, $H_{\ast}(\mathcal{T}(t))=H_{\ast}(\mathcal{E})$, $t>T_{\min}$ (for
 terminological definitions, see Sec.\ref{sec:met}). The minimum time required for this process, $T_{\min}$,
 approximates the physiological \textit{learning time} \cite{PLoS,Rev,Arai,Bso,Hof,CAs,MWind2,Eff,Rev,Kang}.

Second, using a simplicial schema to describe neuronal activity aids in interpreting neurophysiological 
computations. For example, the assemblies igniting along a navigated path, $\gamma$, induce a series of 
consecutively activating simplexes that form a representation of $\gamma$---a simplicial trajectory, 
\begin{equation}
	\Gamma = \{\sigma_1, \sigma_2, \ldots, \sigma_n\},
	\label{Gamma} 
\end{equation}
embedded in $\mathcal{T}$. Such trajectories not only represent ongoing behavior \cite{Brown1,Jensen1,Guger},
but they also enable the reconstruction of an animal's past navigational experiences and the prediction of
upcoming, planned journeys \cite{Fortin1}. Even the possibility of associating neuronal spiking with place
fields relies on the topology of the coactivity complex \cite{Repr,TancerSur}.

The structure of simplicial complexes is derived from the adjacencies of their simplexes. Algorithmically,
this is achieved by introducing an operator, $\partial$, which splits the boundary of each simplex 
(\ref{simplex}) into facets, as illustrated in Fig.~\ref{f1:simp}B,
\begin{equation}
	\partial \sigma_{i_0 \ldots i_k} = \sum_{l=0}^{k}(-1)^{l}\sigma_{i_0 \ldots \cancel{i}_l \ldots i_k},
	\label{dsmplx}
\end{equation}
where the crossed-out index $i_l$ denotes the omitted vertex $v_{i_l}$ \cite{Alexandrov,EdelHarer,ZmBook}.
The key objects in the theory are simplex chains---arbitrary collections of simplexes that span the complex
and collectively capture its shape. For instance, simplicial trajectories (\ref{Gamma}) can be viewed as 
contiguous chains of abutting simplexes.

If a simplicial chain borders a simplex of higher dimensionality, its segments can be ``snapped over" the
boundary of that simplex, producing a chain deformation (Fig.\ref{f1:simp}D). Correspondingly, if two 
chains, $c_1$ and $c_2$, can be deformed into one another---that is, if they differ by a collection of full 
boundaries---they may be considered equivalent, or homologous (Sec.\ref{sec:met}). Simplicial topology exploits
the phenomenon that sets of homologous cycles correspond to elements of the complex's topological structure,
such as holes, cavities, pieces, tunnels, and so forth \cite{Alexandrov,ZmBook,EdelHarer}. Specifically,
classes of equivalent simplicial trajectories (\ref{Gamma}) in representable coactivity complexes mirror the
structure of the environment covered by the place fields, including areas, obstacles encircled by the rat, 
and potential shortcuts the rat can take \cite{PLoS,Arai,Bso,Hof}.

Algebraically, sets of homologous chains form abelian groups or vector spaces, depending on the coefficients
used to count the simplexes. With coefficients from an algebraic field, $\mathbb{K}$, one obtains $n$ vector
spaces, one for each dimension: $H_0(\mathcal{T},\mathbb{K}),H_1(\mathcal{T},\mathbb{K}),\ldots,H_n(\mathcal{T},
\mathbb{K})$, commonly referred to as the simplicial homologies of $\mathcal{T}$. The dimensionalities of these
homology spaces, $b_k = \dim(H_k(X, \mathbb{K}))$, known as Betti numbers, count connectivity components 
($\beta_0$), holes and tunnels ($\beta_1$), cavities ($\beta_2$), and so on. In this study, we use the binary
field, $\mathbb{K} = \mathbb{Z}_2 =\{0, 1\}$, which is used without explicit reference.
%%%%%%%%%%%%%%%%%%%%%%%%%%%%%%%%%%%%%%%%%%%%%%%%%%%%%%%%%%%%

%\begin{figure}[h]
%\end{figure}

%%%%%%%%%%%%%%%%%%%%%%%%%%%%%%%%%%%%%%%%%%%%%%%%%%%%%%%%%%%% 
In practice, the hippocampal coactivity complex is often constructed as the clique complex associated with the
graph of coactivity, $\mathcal{G}$, also known as the cognitive graph \cite{Chras,Peer,cGr,Burgess,Trullier,
	Eriq,Farzanfar}. The nodes of this graph correspond to principal cells, and its links represent either
physiological or functional connections between them. If these connections are directed (e.g., extending from
presynaptic to postsynaptic cells), the graph is directed and referred to as a \textit{digraph}. Otherwise, if
the connections represent undirected coupling, such as the rate of coactivity between cells, the coactivity 
graph is undirected.

In applications, specific constructions of cliques may be used to capture different physiological phenomena.
For instance, cliques formed over time from frequently co-occurring lower-order coactivities (e.g., pairs or
triples of cells) can model input-integrating cell assemblies, whereas large groups of simultaneously co-firing
cells can represent coincidence-detecting cell assemblies. Both mechanisms have been experimentally identified,
making the corresponding coactivity complexes viable functional representations of physiological coactivity
pattern \cite{Arai,Bso,Hof,CAs,MWind2,Eff,Rev,Kang}.

\subsection{Sequential schema of neuronal activity}
\begin{figure}[h]
	%\begin{wrapfigure}{c}{0.75\textwidth}
	\centering
	\includegraphics[scale=0.78]{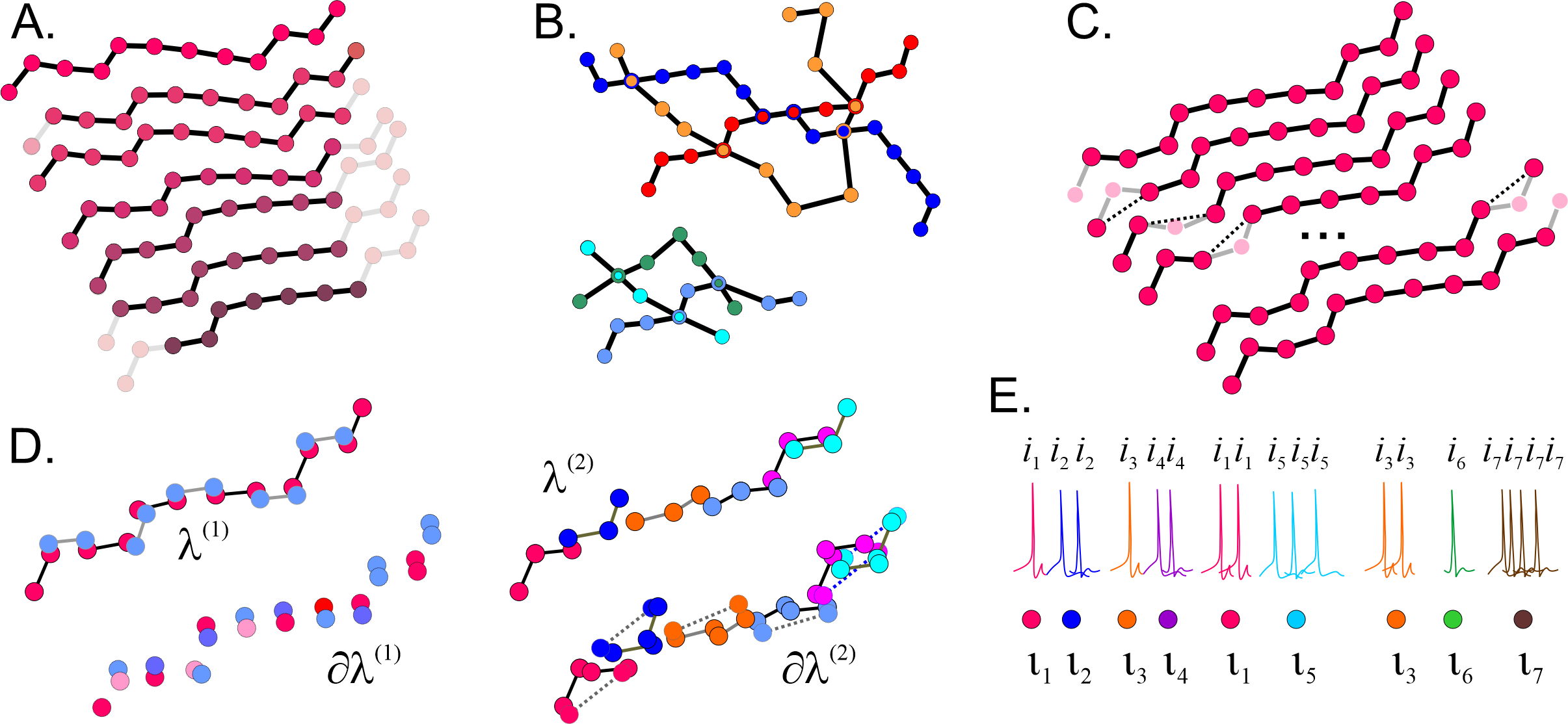}
	\caption{\footnotesize{\textbf{Path, path chains, path complexes}. \textbf{A}. An elementary path $e$ of
			length $12$ and examples of shorter paths obtained by 'plucking' its end vertexes. All truncated
			paths must be included into each path complex that contains $e$.
			\textbf{B}. A path complex with two-components and six maximal elementary paths. Presence of the
			subpaths obtained by ``plucking" the longest paths is implied.
			\textbf{C}. The path-boundary (\ref{dpath}) of the sequence $e$ illustrated on the top of panel A.
			Note that only the first and the last paths shown on this panel also appear on panel A.
			\textbf{D}. A sequence of $k$ elements can be viewed as combination of pairwise links (length-one
			path chain, $\lambda^{(1)}$, left panel), or triples of consecutive vertexes (length-two path chain,
			$\lambda^{(2)}$, right panel), etc. The boundaries of such chains, $\partial\lambda^{(1)}$ and 
			$\partial \lambda^{(2)}$, are shown below.
			\textbf{E}. A collection of bursting neurons, $\iota_1,\iota_2,\ldots,\iota_7$, produce series of
			spikes marked by repeating indexes, $i_k$: neuron $\iota_2$ yields two spikes in a burst, neuron
			$\iota_5$ yields three spikes, neuron $\iota_7$ yields four. Regularized sequence reveals the order
			in which the neurons fire (bottom).
	}}
	\label{f2:de}
	%\end{wrapfigure}
\end{figure}

If ordered sequences of neurons' firings, rather than nearly-independently igniting cell groups, serve as
functional units of neuronal activity, then the approach described above requires principal modifications.
Let us represent the population of active units (cells or their assemblies) by a set of vertexes, $V=\{v_1,
	v_2, \ldots, v_n\}$, and consider sequences of various lengths,
\begin{equation}
	e_i=\{v_{i_0},v_{i_1}, \ldots,v_{i_k}\}\equiv e_{i_0i_1\ldots i_k}. 
	\label{esmplx}
\end{equation}
Following the terminology of path homology theory, we will also refer to (\ref{esmplx}) as an 
\textit{elementary path} running through the set $V$ \cite{Grig1,Grig2}. A simplicial path (\ref{Gamma})
is one example of an elementary path comprised of igniting cell assemblies, while a sequence of individual
neurons activated autonomously in the animal's brain (e.g., during sleep) is another \cite{Olaf,Farooq,Dragoi3}.
Note that, since the firing units in (\ref{esmplx}) spike in order at times $t_{i_0}<t_{i_1}<\ldots < t_{i_k}$,
each sequence (\ref{esmplx}) is characterized by a specific start and completion time. The set of spiking 
sequences produced by a particular moment $t$ then forms a \textit{path complex},
\begin{equation}
	\mathcal{P}(t)=\cup_{t_i<t}\,e_{i_0i_1\ldots i_k},
	\label{pcomplex}
\end{equation}
which is a path analogue of $\mathcal{T}(t)$ (Fig.~\ref{f2:de}A). The only requirement for the collection of
paths (\ref{pcomplex}) is that if a path $e_i$ is included, then the paths obtained by ``plucking" the ends
of $e_i$ must also belong to $\mathcal{P}(t)$ (Fig.~\ref{f2:de}B). In other words, given a cell sequence 
(\ref{esmplx}), its shorter contiguous subsequences are assumed to contribute to the informational framework
encoded by $\mathcal{P}(t)$---a physiologically natural assumption.

Note that building a path complex, \textit{e.g.}, constructing a path complex from a dataset of spiking series,
requires specifying all elementary paths: the existence or absence of paths should not be presumed.
However, if sequences follow a particular grammar, it may be possible to identify a specific ``template" that
generates them \cite{Farooq,Dragoi3,Vaz}. For instance, if a complex $\mathcal{P}$ contains paths assembled from
specific pairs of vertices, then $\mathcal{P}$ can be induced by traversing a certain graph $G$, where these 
selected pairs are connected by edges \cite{Grig1,Grig2}. In such a case, given a finite set of observed paths,
one may assume that the observations are partial, while the ``ultimate" path complex encompasses all paths
consistent with the graph's connectivity. This allows for the study of the pool of all possible vertex sequences.
Physiologically, such a graph may emerge, \textit{e.g.}, as the synaptic configuration of the underlying network,
or as a coactivity graph, $\mathcal{G}$ \cite{Vaz}. In the following sections, we will focus on these 
\textit{graph-representable} path complexes, although the theory allows for more general constructions.
%%%%%%%%%%%%%%%%%%%%%%%%%%%%%%%%%%%%%%%%%%%%%%%%%%%%%%%%%%%%

%%%%%%%%%%%%%%%%%%%%%%%%%%%%%%%%%%%%%%%%%%%%%%%%%%%%%%%%%%%%
The homological description of path complexes develops similarly to simplicial homology theory. Viewing an
elementary path (\ref{esmplx}) of length $k$ as analogous to a $k$-dimensional simplex, one can construct its
``structural boundary" as the formal alternating sum, 
\begin{equation}
	\partial e_{i_0\ldots i_k}=\sum \limits_{l=0}^{k}(-1)^{l}e_{i_0\ldots\cancel{i}_l\ldots i_k}, 
	\label{dpath}
\end{equation}
where $e_{i_0\ldots\cancel{i}_l\ldots i_k}$ runs from $v{i_0}$ to $v_{i_k}$, omitting the vertex $v_{i_l}$
(Fig.~\ref{f2:de}C). Thus, the ``boundary" of a path of length $k$ is a combination of its $(k+1)$ subpaths,
each of length $k-1$.
	
The next step is to consider \textit{path chains}---arbitrary combinations of elementary paths---and to view two
chains as homologous to one another if their difference is the boundary of another path chain, in the sense of
formula (\ref{dpath}) \cite{Grig1,Grig2}. The resulting equivalence classes, properly counted, form vector
spaces called the \textit{path homologies} of a path complex $\mathcal{P}$, denoted below as $\HP_0(\mathcal{P}),
\HP_1(\mathcal{P}), \ldots$, which describe its structure just as simplicial homologies describe simplicial
complexes.

This outline requires several stipulations. First, unlike the subsimplexes on the right-hand side of
(\ref{dsmplx}), which architecturally belong to $\mathcal{T}(t)$, the ``path-facets" on the right-hand side
of (\ref{dpath}) may not fit the path complex, meaning that formula (\ref{dpath}) may be ill-defined. For
example, consider a path complex $\mathcal{P}_{e}$ that contains a single linear path (\ref{esmplx}) and
its required ``truncations." In this case, formula (\ref{dpath}) generates subpaths that skip intermediate
vertices and therefore do not belong to $\mathcal{P}_{e}$ (Fig.~\ref{f2:de}C).

This perplexity is resolvable (see below) and even desirable: if paths responded to the boundary operator
(\ref{dpath}) in the same way that simplexes respond to the boundary decomposition (\ref{dsmplx}), the two
homology theories would simply emulate one another. Instead, the principles for identifying structurally
equivalent serial patterns of activity are genuinely different from simplicial equivalence, pointing to 
qualitatively distinct organizations of information. 

Of course, paths that behave like simplexes do exist, but they are non-generic and consist only of those that
traverse simplex vertices. A complex $\mathcal{P}$ comprised exclusively of such ``folded-into-a-simplex" paths,
which contain \textit{all} boundary subpaths of \textit{all} of its elementary paths, is structurally similar
to a simplicial complex and is referred to as perfect in \cite{Grig1,Grig2,Grig3,Grig4}. However, most path 
complexes are far from perfect; that is, boundary subpaths of most of their elementary paths are missing, 
resulting in path homologies that differ substantially from the simplicial homologies of the underlying graph's
clique complex.

The second point is that there are multiple ways to interpret a given firing series as a path. A string of
$k$ vertices can be viewed not only as a single path of length $k$, but also as a combination of $1$-vertex 
paths---individual nodes (a $0$-chain, $\lambda^{(0)}$), a collection of $2$-vertex paths (a $1$-chain of
graph links, $\lambda^{(1)}$), or as sequences of $3$-vertex paths, \textit{i.e.}, link pairs (a $2$-chain,
$\lambda^{(2)}$), link triples (a $3$-chain, $\lambda^{(3)}$), and so on, with all shorter constituents 
overlapping in arbitrary ways (Fig.~\ref{f2:de}D). As it turns out, reasoning in terms of such ``path chains,"
rather than individual paths, allows for rectifying the unencumbered boundary operator (\ref{dpath}) and 
transforms any collection of paths into a self-contained complex \cite{Grig1,Grig2,Grig3,Grig4}.

Third, paths that repeatedly traverse the same vertices, such as $e_{i_1i_2i_2i_3i_4i_4i_1i_1i_5i_5i_5\ldots}$,
are possible but often redundant in practical analyses. For example, consider a complex where the elementary 
paths correspond to sequences of spikes, each indexed by the neuron that produced it. Since neurons tend to 
``burst," firing spikes in rapid succession, such paths will contain multiple repeating indices (Fig.\ref{f2:de}E,
\cite{Bertram,Izh1}). In analyses focusing on the order of neuron activation, these repetitions are uninformative
and should be factored out. Conveniently, path homology theory allows for the exclusion of index repetitions by
reducing each path to its unique \textit{regular representative} with distinct consecutive vertices \cite{Grig1,
	Grig2}. At the graph level, this corresponds to removing trivial loops, $e_{ii}$. The resulting paths form
\textit{regular path complexes}, described by \textit{regular path homologies}. Below we consider only these 
regular path complexes and omit the term ``regular" (see Sec.\ref{sec:met}).

With these provisions, path homologies provide a natural, context-independent description of serial activity
pools, and, together with simplicial homologies highlight principles by which neuronal activity may be organized.
In the following sections, we explore this question by simulating the simplicial complexes of coactivity and the
complexes of firing sequences produced by the hippocampus during spatial navigation. We then evaluate their 
spatial and ordinal structure and discuss certain physiological implications of their homological classification.

\section{Results}
\label{sec:res}

\begin{figure}[h]
	%%%\begin{wrapfigure}{c}{0.5\textwidth}
	\centering 
	\includegraphics[scale=0.78]{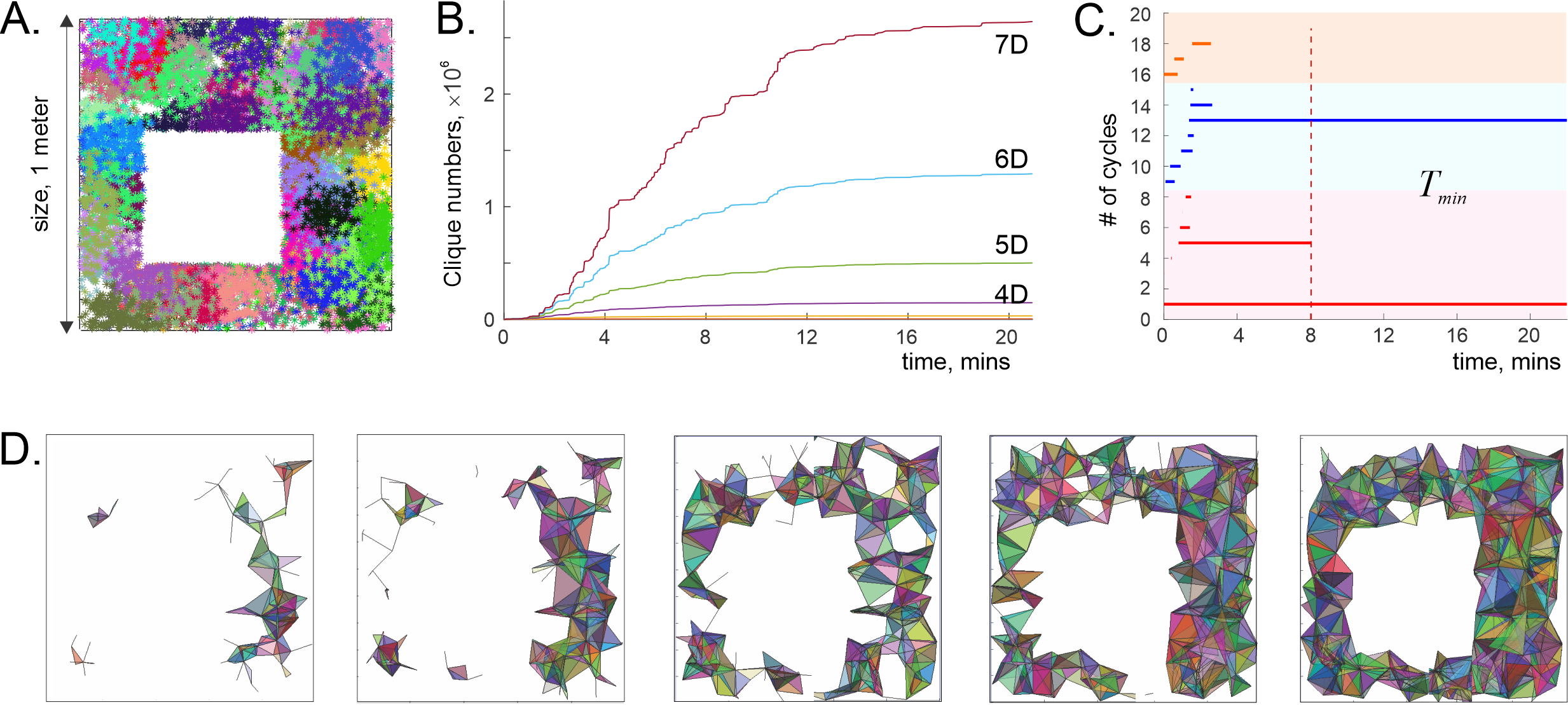}
	\caption{{\footnotesize
			\textbf{Simplicial dynamics}.
			\textbf{A}. $2D$ arena with one hole, $\mathcal{E}^{(2)}_1$, sized $1\textit{m}\times1\textit{m}$,
			typical for electrophysiological experiments \cite{Hafting,BrunG}, covered by $200$ simulated place
			fields. The cells fired in the order their fields were traversed. Each dot of a particular color 
			represents a spike fired by the	corresponding cell at that location. Each cluster of colored dots
			represents a place field. 	
			\textbf{B}. As the animal navigates, the set of cells that have spiked grows, as does the pool of
			coactive cell combinations. Shown is the growing population of seven-dimensional ($7D$), 
			six-dimensional ($6D$), five-dimensional ($5D$), \textit{etc}., cliques of $\mathcal{G}$, \textit{i.e.},
			simplexes of $\mathcal{T}(\mathcal{G})$.
			\textbf{C}. The corresponding topological evolution of the coactivity complex: at first, 
			$\mathcal{T}$ is small and fragmented, but as the complex grows in size, its structure simplifies.
			Top, orange streaks mark the timelines of the noncontractible $2D$ cavities, middle, blue timelines
			pertain to $1D$ holes and the bottom, red timelines follow $0D$ loops---pieces of $\mathcal{T}$.
			From early on, higher-dimensional Betti numbers, $b_{D\geq2}$, are suppressed, allowing 
			representability of the complex \cite{Repr}. The spurious holes close up by $T_{\min}=3.2$ min,	and
			complex fuses into one piece in about $8.0$ mins, at which point $\mathcal{T}$ becomes topologically
			equivalent to $\mathcal{E}^{(2)}_1$.
			\textbf{D}. Pictorial dynamics of the coactivity complex, with $2D$ facets projected into the 
			navigated environment.
	}}
	\label{f3:env}	
	%%%\end{wrapfigure}
\end{figure}

We simulated the rat's movement in a small, low-dimensional enclosure covered by randomly scattered place
fields (Fig.\ref{f3:env}A). From the resulting spiking patterns, we constructed the coactivity graph,
$\mathcal{G}$, by identifying pairs of frequently cofiring cells \cite{cGr,PLoS,Arai,Bso}. The corresponding
coactivity complex is dynamic (Fig.\ref{f3:env}B): in the early stages of navigation, when only a few cell 
groups had a chance to cofire, $\mathcal{T}$ is small, fragmented, and contains numerous holes, cavities, 
tunnels, and other features, most of which lack physical counterparts. When the spiking parameters are 
biologically realistic, these ``spurious" structures tend to disappear as the pool of coactivities grows
(Fig.~\ref{f3:env}C). Consequently, the complex $\mathcal{T}$ becomes representable and takes on the 
topological shape of the surrounding environment \cite{Repr}.

In other words, as spiking accumulates, the pool of homologous simplicial chains consolidates, revealing the
topological structure of the underlying space (Fig.~\ref{f3:env}B) \cite{PLoS,Arai,Hof,Bso,MWind2,Eff,CAs,Rev,
	Kang}, which is reflected in the animal's spatial behavior \cite{Poucet,Alv1,Goodrich,Chen,Wu,Alv2,Alv3}.
%%%%%%%%%%%%%%%%%%%%%%%%%%%%%%%%%%%%%%%%

%%%%%%%%%%%%%%%%%%%%%%%%%%%%%%%%%
On the other hand, the very same process can be interpreted as the formation of a path complex of simplicial
trajectories. Indeed, the simplicial complex $\mathcal{T}$ may be viewed as one or several ``folded" simplicial
paths: the assemblies that ignite at the animal's current position produce active simplexes at the ``tip" of an
unfolding simplicial path $\Gamma$, while the simplexes left behind in its ``tail" accumulate into $\mathcal{T}$.

Furthermore, at each moment of its development, $\mathcal{T}(t)$ can be represented by the maximal simplex
connectivity graph, $\mathfrak{G}_{\mathcal{T}}(t)$, whose vertices correspond to the cell assemblies---the
maximal simplexes $\sigma_i\in\mathcal{T}(t)$---connected by an edge $e_{ij}$ if $\sigma_i$ overlaps with 
$\sigma_j$. The simplicial paths $\Gamma$ then correspond to sequences of edges in $\mathfrak{G}_{\mathcal{T}}$.
In the spirit of the persistence approach \cite{Wei,Chaplin2}, the path homologies of this maturing graph reveal,
moment by moment, how the ordinal structure of igniting cell assembly series unfolds over time. This represents
a new kind of homological dynamic that may also influence the animal's behavior and cognitive performance.

For the neuronal ensemble illustrated in Fig.~\ref{f3:env}C, the graph $\mathfrak{G}_{\mathcal{T}}$ initially
appears fragmented, as indicated by large values of the path-homological $0^\textrm{th}$-order Betti numbers,
$\betaup_0(\mathfrak{G}_{\mathcal{T}})\sim 8$. This is expected, as the disconnected subgraphs correspond to 
the connected components of $\mathcal{T}$; the $0^{\textrm{th}}$ order topological loops appear and disappear
simultaneously, indicating the consolidation of spatial and ordinal coactivity domains, $\beta_0(\mathcal{T})
= \betaup_0(\mathfrak{G}_{\mathcal{T}})$.
%%%%%%%%%%%%%%%%%%%%%%%%%%%%%%%%%%%%%%%%

\begin{figure}[h]
	%\begin{wrapfigure}{c}{0.65\textwidth}
	\centering 
	\includegraphics[scale=0.8]{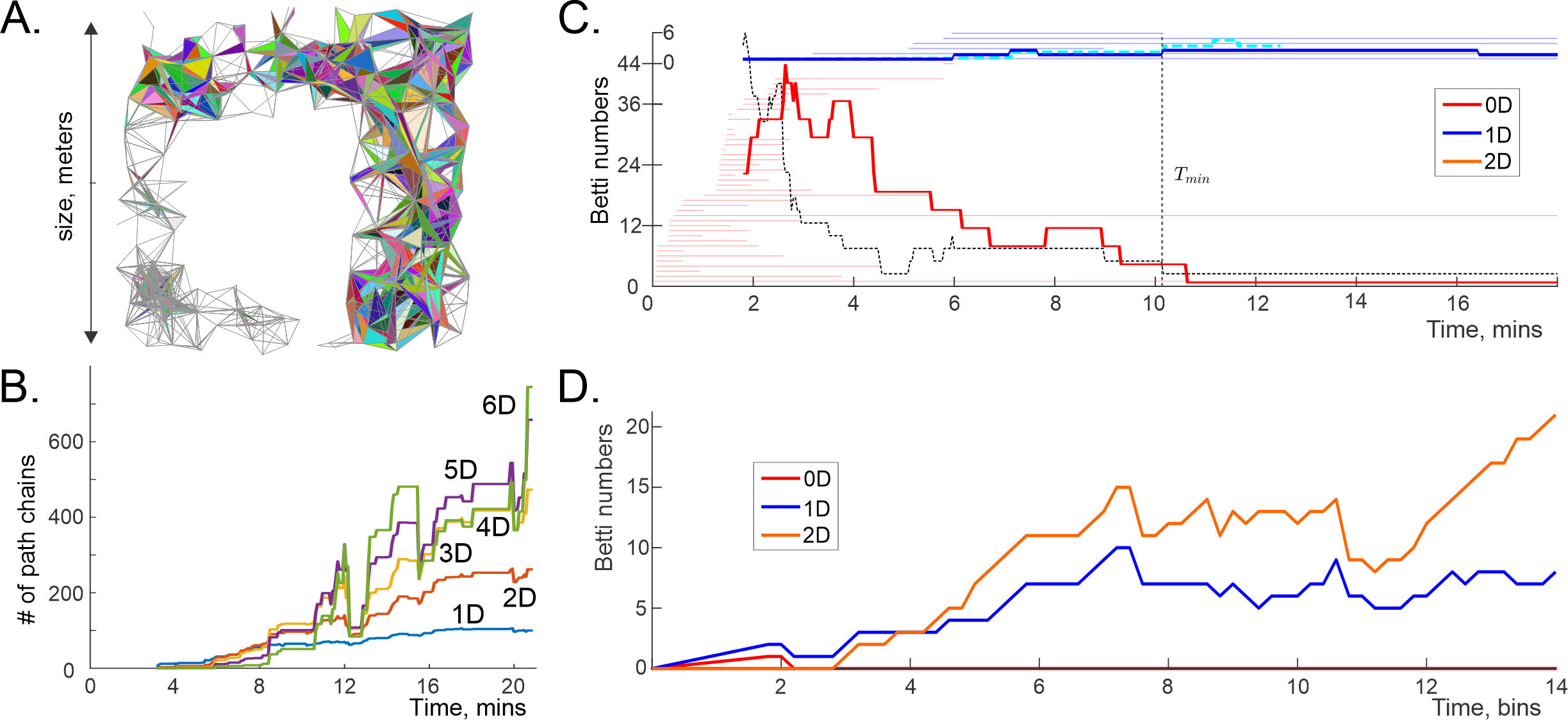}
	\caption{\footnotesize{\textbf{Temporal dynamics of cell assembly sequences}. 
		\textbf{A}. Illustrative representation of the coactivity complex's connectivity graph $\mathfrak{G}
		_{\mathcal{T}}$ (gray bonds), which defines incidences between the maximal simplexes in $\mathcal{T}
		_{\mathcal{G}}$ at a particular stage of its development.
		\textbf{B}. The population of $5$-vertex, $4$-vertex, $3$-vertex, ..., path-chains in the path complex
		$\mathcal{P}_{\mathfrak{G}}$, induced by the maximal simplex connectivity graph $\mathfrak{G}_
		{\mathcal{T}}$, grows over time.
		\textbf{C}. The path-homological dynamics of $\mathcal{P}_{\mathfrak{G}}$ differ qualitatively, 
		exhibiting similarities with the simplicial-homological dynamics: the number of distinct path-cycles
		initially increases and then decreases over time. The plot shows the numbers of $0D$ (blue, 
		$\betaup_0$), $1D$ (red, $\betaup_1$), and $2D$ (orange, $\betaup_2$) path-cycles. Higher-order path
		chains consolidate into a single homology class ($\betaup_{k>2} = 0$). The horizontal lines in the
		background indicate the timelines of spurious topological loops in $\mathcal{T}_{\mathcal{G}}$, 
		identified by simplicial persistent-homological analyses (see Fig.~\ref{f3:env}C, \cite{PLoS}).
		\textbf{D}. The number of path-homologically distinct path-cycles in a growing random graph
		(connected, $\betaup_0=1$) increases, whereas the number of heterologous chains also remains high
		($D\gtrsim 6$, not shown).
	}}
	\label{f4:persmx}	
	%\end{wrapfigure}
\end{figure}

%%%%%%%%%%%%%%%%%%%%%%%%%%%%%%%%%%%%%%%
On the other hand, first-order homologies, $H_1(\mathcal{T})$ and $\HP_1(\mathfrak{G}_{\mathcal{T}})$, may
differ significantly from each other. This is because $1D$ link-chains in $\mathcal{T}$ ``cycle" relative to
their simplex boundaries (\ref{dsmplx}) differently than the chains of $\mathfrak{G}_{\mathcal{T}}$ edges 
``cycle" relative to the path boundaries (\ref{dpath}). In particular, the high computational cost of 
evaluating path homologies (Fig.\ref{f4:persmx}B) limits the sizes and the densities of the coactivity graphs
that can be used in constructing path complexes, \textit{i.e.}, the sizes of spiking populations. Therefore,
we restricted our analyses to the series of the most prominently firing cells---those that produce at least
four spikes per burst (Fig.\ref{f2:de}E, \cite{Bertram,Izh1}).

The results show that the homological dynamics of the emerging path complexes, $\mathcal{P}_{\mathfrak{G}}(t)$,
share several similarities with the dynamics of the coactivity complex $\mathcal{T}_{\mathcal{G}}(t)$: the 
population of distinct $1D$ cycles initially increases to high values (with the bursting threshold $n_b\geq3$),
then decreases and stabilizes as distinct serial patterns become path-homologous. This happens over a period
$T_{\min}^{(p)}$ that slightly exceeds the simplicial learning time. 
A similar trend is observed in the population of $2D$ loops, while higher-order Betti numbers, $\betaup_{k>2}$,
tend to vanish. The latter indicates that homologically distinct chains composed of long ($k > 3$) elementary
paths tend to consolidate. In other words, path-homology analysis suggests that longer sequences of assemblies
ignited by traversing firing field maps tend to be homologous to one another, thus capturing less information,
whereas collections of shorter sequences form richer structures.

Importantly, this phenomenon is not generic: in our simulations, path-homological Betti numbers of random graphs
tend to increase with the graph's size \cite{Chaplin1}, suggesting that the ``homological equalizing" of longer
cell assembly sequences may be a property of map-representing graphs. 

Adding directionalities to the edges of $\mathfrak{G}_{\mathcal{T}}$, \textit{i.e.}, simulating preferred 
directions of activity propagation, does not qualitatively alter the previous scenario. The dynamics of 
$\betaup_0$ remains unchanged, and $\betaup_1$ shows exuberant dynamics at the early stages of navigation,
which then subsides (longer timescales, with lower intermediate values).

\begin{figure}[h]
	%\begin{wrapfigure}{c}{0.65\textwidth}
	\centering 
	\includegraphics[scale=0.79]{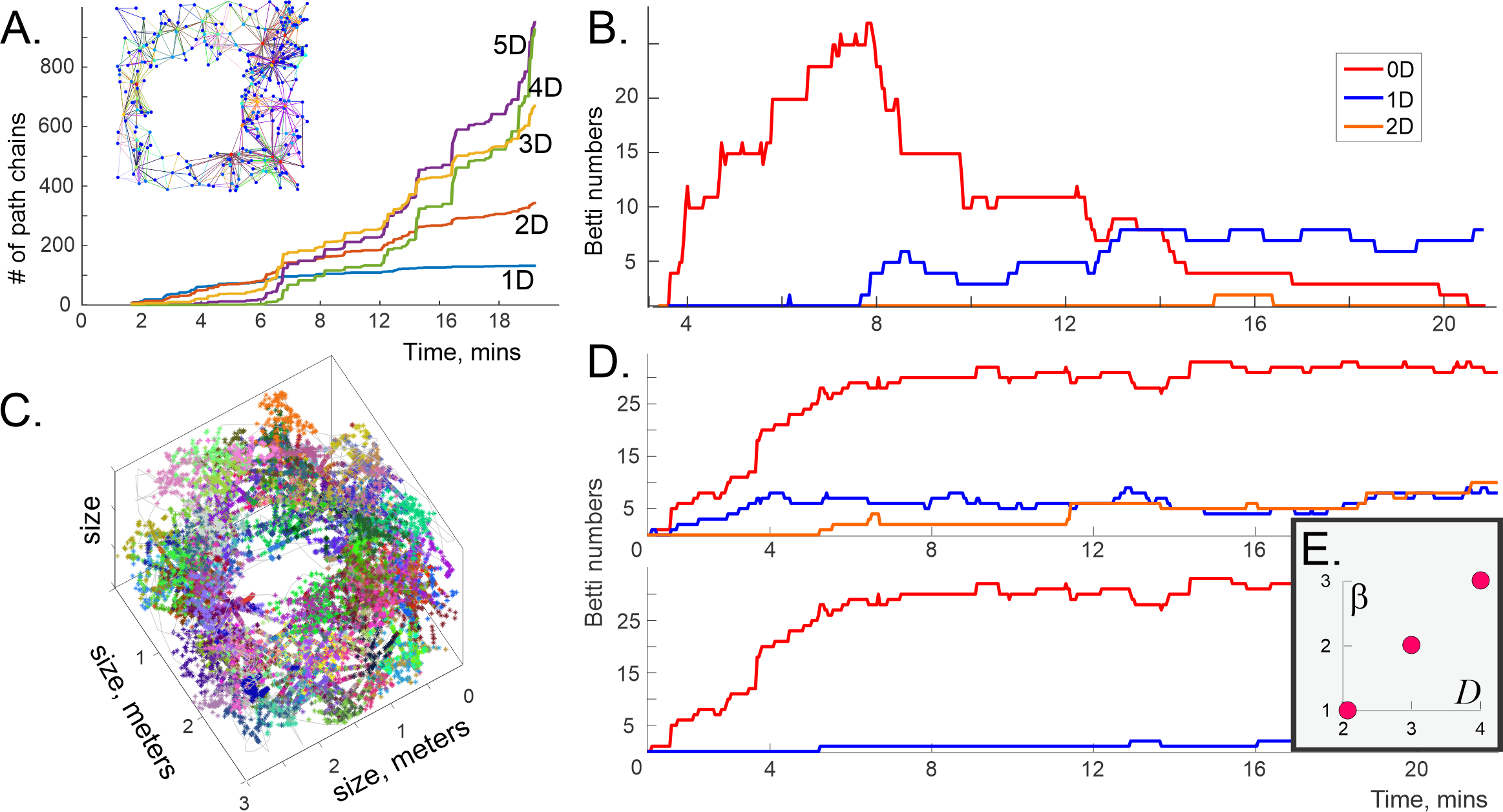}
	\caption{{\footnotesize 
			\textbf{Temporal dynamics}. \textbf{A}. The population of $5D$, $4D$, $3D$, ... path-chains in the
			path complex induced by the growing cognitive graph $\mathcal{G}$ (embedded panel) grows with time.
			\textbf{B}. Path-homological dynamics of the coactivity graph $\mathcal{G}$ (neuronal sequences) is
			consistent with the dynamics of $\mathcal{P}_{\mathfrak{G}}$ (cell assembly sequences): the number
			of distinct path cycles (Betti numbers $\betaup_k$) first grows and then decreases with time. 
			Higher-order path cycles vanish.
			\textbf{C}. Place field map in a $3D$ environment with one hole, $\mathcal{E}^{(3)}_1$ (simulated 
			bat movements \cite{Hof}). Points correspond to	spikes fired by a specific cell at a particular
			location.
			\textbf{D}. $3D$ navigation produces more path-homologically distinct length-$0$ path cycles, which
			means that for the select size of place cell ensemble, speed value and the spiking parameters, the
			coactivity graphs (undirected, top and directed, bottom) have many disconnected components. Note 
			also that the undirected coactivity graph generates a population of persistent, homologically 
			distinct length-$2$ cycles, which appear only fleetingly in panel B.
			\textbf{E}. Lasting ($|T|\gtrsim T_{\min}$) path-homological Betti-order of the undirected cell
			assembly connectivity graph grows with the dimensionality of the navigated space.
	}}
	\label{f5:persgr}	
	%\end{wrapfigure}
\end{figure}

\textbf{Sequences of neuronal activity} traced in most electrophysiological studies follow the connectivity of
the original cognitive graph $\mathcal{G}$, where vertices correspond to individual cells and links represent
functional or synaptic connections between them. This graph is typically sparser than $\mathfrak{G}_{\mathcal
	{T}}$, which relaxes the numerical restrictions on firing activity. We simulated the development of this
graph during the rat's running session discussed above (Fig.~\ref{f3:env}A), focusing on cells that produced at
least three spikes per burst (Fig.~\ref{f5:persgr}A). The resulting path-homological dynamics exhibited familiar
characteristics: initially, $\mathcal{G}$ induces many spurious, heterologous sequences, which is followed by a
period of consolidation and stabilization (Fig.~\ref{f5:persgr}B).

The fact that this period aligns closely with the cell assembly sequence learning timescale suggests that
information encoded by both the neuronal and cell assembly sequences becomes accessible over similar timeframes.
Additionally, the number of homologically distinct, closed neuronal firing sequences is comparable to the number
of heterologous cell assembly path cycles and the number of simplicial loops in $\mathcal{T}$ (Figs.~\ref{f3:env},
\ref{f4:persmx}). If the connectivity graph is directed, reflecting, \textit{e.g.}, the synaptic connections
between neurons, then the intermediate population of nonidentical short sequences is higher ($\betaup_1$ and 
$\betaup_2$ grow), but the overall path-homological dynamics remains similar.

In a $3D$ environment, the connectivity graphs induced from the place field maps produce a population of 
homologically distinct $2D$ path-loops, in both vertex (neurons) and clique (cell assembly) sequences, that
persist through the simulated period. To obtain these results, we simulated moves through a $3D$ topological
cylinder over the $2D$ environment shown on Fig.~\ref{f3:env}A, $\mathcal{E}^{(3)}_1=\mathcal{E}^{(2)}_1\times
I$, where $I=[0,L]$ is a Euclidean interval (Fig.~\ref{f5:persgr}C). The size of the base and the height, $L$,
were scaled to match the proportions of a bat's cave described in \cite{Yarts}, about $3$ meters across
\cite{Hof}. The resulting space has the same fundamental group, $\pi_1(\mathcal{E}^{(3)}_1)=\pi_1(\mathcal{E}
^{(2)}_1)=\mathbb{Z}$, and the same topological complexity \cite{Schwarz,Farber}, and therefore poses a 
comparable topological learning task as its $2D$ counterpart. The lengthening of the path-homologically
nontrivial spiking chains may hence be attributed to the increase of the underlying space's dimensionality
(Fig.~\ref{f5:persgr}D).

To investigate whether the maximal length of path-homologically distinct sequences,
\begin{equation}
	\betaup_{\max}=\max_{k,t}(\{\betaup_{k}(t)>0\}),
	\label{ler}
\end{equation} 
and its duration, $|T|$, where
\begin{equation}
	T=\{t: \max_{k}\betaup_k(t)=\betaup_{\max}\},
\end{equation}
continue to increase with the dimensionality of the underlying space, $D=\dim(\mathcal{E})$---a dynamic also
observed in the simplicial homologies of the coactivity complex \cite{Repr}---we leveraged the fact that all
components of our construction (trajectory, place fields, etc.) can be extended to any dimension without 
affecting the topological complexity or the homotopical and simplicial-homological structure of the space.

Specifically, we constructed several coactivity graphs from the place field maps covering a $4D$ analogue of
the $3D$ cave (Fig.~\ref{f5:persgr}C) and $2D$ navigational arena (Fig.~\ref{f3:env}A), $\mathcal{E}^{(4)}_1=
\mathcal{E}^{(3)}_1\times I=\mathcal{E}^{(2)}_1\times I\times I$, and observed that the \textit{path-homological
Leray index} (\ref{ler}) increased to $\betaup_4=3$. The emerging $\betaup_D\propto D$ trend in map-induced
path complexes (Fig.~\ref{f5:persgr}E), as well as the general suppression of higher-order Betti numbers 
requires further investigations. Biologically, this would imply that longer sequences tend to homogenize,
whereas collections of shorter sequences exhibit greater structural richness. 
%%%%%%%%%%%%%%%%%%%%%%%%%%%%%%%%%%%%%%%%

%%%%%%%%%%%%%%%%%%%%%%%%%%%%%%%%%%%%%%%
\subsection{Memory spaces}
Episodic memory frameworks can be regarded as spaces in which specific memories, $m_1$, $m_2$,..., correspond
to regions, $r_1, r_2, \ldots$ \cite{Eich1}. Neurophysiologically, this perspective is instrumental because it
allows viewing many cognitive phenomena---such as spatial planning and exploration, transitive and relational
inference, memory retrieval, and others---as particular cases of ``mental navigation," thereby facilitating 
physiological interpretations of data \cite{Eich1,Eich2,Viewpoi,Crystal,Deuker,Liu2,SchemaM}.

From a topological standpoint, this interpretation is also natural, as simplicial complexes define finitary
spaces: individual simplexes emulate points, $x_i$ (``memory nodes"), and their agglomerates represent broader
memory scopes, $r_i$. Essentially, this construction extends the cell assembly theory \cite{Nicl,Tzl},
which associates cognitive episodes with the activity of cell assemblies. It posits mappings from coactivity
simplexes or their complexes to a cognitive map or memory space, 
\begin{equation*}
	f:\sigma_i\to m_i\in\mathcal{M}.
\end{equation*} 
The totality of such mappings defines a \textit{singular coactivity complex} associated with $\mathcal{M}$,
whose homologies determine its topological structure \cite{SchemaM}.

As mentioned above, a remarkable property of hippocampal coactivity complexes is that they are 
\textit{representable}, meaning there exists a domain in a low-dimensional Euclidean space, $X$, covered by
a set of regular regions, $\upsilon_1, \upsilon_2, \ldots, \upsilon_N$, whose nonempty overlaps correspond
one-to-one with the coactivity simplexes. Specifically, $\upsilon_{i_1} \cap \upsilon_{i_2} \cap \ldots \cap
\upsilon_{i_k} \neq \varnothing \implies \sigma_{i_1 i_2 \ldots i_k} \neq \varnothing$ \cite{Repr,TancerSur}.
As discussed in \cite{Repr}, it is the dimensionality of $X$ that limits the order of non-vanishing Betti 
numbers, $\beta$s. The experimental discovery of hippocampal representability---the existence of place fields
marked a major advance in our understanding of learning, memory, and cognition. In particular, the match 
between \v{C}ech homologies of the place field maps and the simplicial homologies of the coactivity complex
supports interpreting the latter as a discrete topological map of the navigated space \cite{OKeefe,KropffV,
	Grieves}. Similarly, the match between the simplicial homologies of place cell coactivity and the singular
homologies of the corresponding discrete Alexandrov space suggests viewing the latter as a discretized
representation of the environment embedded within memory space $\mathcal{M}$ (Fig.~\ref{f3:env}A)
\cite{McCord,Raume,Fth,Stng,Osk}.

The possibility of describing the ordinal structure of episodic memories by classifying sequences of regions
distributed over $\mathcal{M}$,
\begin{equation}
	\wp=\{r_1,r_2,\ldots,r_m\},
	\label{apath}
\end{equation}
opens a complementary avenue for analyses. Structurally, the ``topological fabric" of a modeled memory space is
defined by the pattern of immediate (minimal) neighborhoods, $U_i$, of its points, which is described by the 
\textit{topogenous matrix} \cite{Shiraki1}:
\begin{equation}
	t_{ij}=
	\begin{cases}
		1, \,\,\, & \mbox{if $x_i\in U_j$,} \\ 
		0 \,\,\, & \mbox{otherwise}.
	\end{cases}
	\label{inc}
\end{equation} 
Here, we assume that the memory spaces are $T_0$-separable, meaning that different points, $x_i \neq x_j$,
have distinct minimal neighborhoods, $U_i \neq U_j$, and we set $t_{ii} = 0$ \cite{Raume,Fth,Stng,McCord,Osk}.
The matrix (\ref{inc}) can then be viewed as the connectivity matrix of a \textit{topogenous graph} $\mathscr
{G}$, where vertices represent elementary memory nodes and edges define topological overlaps between them.\footnote{The
	graph $\mathscr{G}$ also describes connectivity between all simplexes in the coactivity complex.} The path
homologies of the graph $\mathscr{G}$ can thus describe the ordinal organization of memory sequences.
%%%%%%%%%%%%%%%%%%%%%%%%%%%%%%%%%%%%%%%%

\begin{figure}[h]
	%\begin{wrapfigure}{c}{0.65\textwidth}
	\centering 
	\includegraphics[scale=0.8]{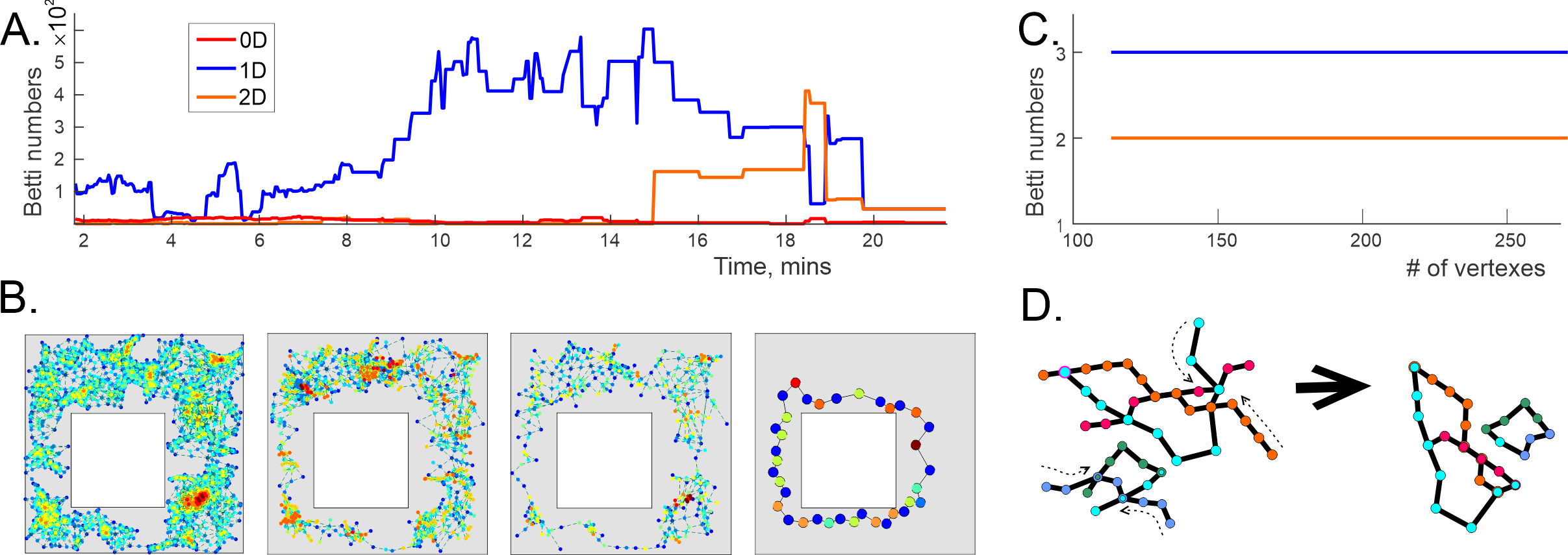}
	\caption{{\footnotesize
			\textbf{Temporal dynamics}.
			\textbf{A}. Path-Betti numbers for an series of topogenous graphs pertaining to the Alexandrov
			spaces induced at different stages of the coactivity complex's development. The topological 
			dynamics, as described by simplicial and singular homologies is identical to the one shown on
			Fig.~\ref{f3:env}A.
			\textbf{B}. Elementary locations of the Alexandrov space built for place field map shown on
			Fig.~\ref{f3:env}A, and the location maps corresponding to a few of its retractions. The 
			rightmost panel shows the core---maximally reduced schema of the memory space, containing only
			$36$ locations. Color represents the dimensionality of the simplexes giving rise to the points
			\cite{CAs,Reimn}.
			\textbf{C}. Path-homological Betti numbers computed for the topogenous graphs of a retracting
			Alexandrov space remain unchanged.
			\textbf{D}. An example of a homotopy transformation of two-component graph that eliminates the
			tentacles. The homotopy of the corresponding path complexes is implied.
	}}
	\label{f6:stng}	
	%\end{wrapfigure} 
\end{figure}

%%%%%%%%%%%%%%%%%%%%%%%%%%%%%%%%%%%%%%%
We constructed the graph $\mathscr{G}$ for the Alexandrov spaces corresponding to the snapshots of the
evolving coactivity complex $\mathcal{T}(t)$, focusing on the most prominently firing cells (at least 
six spikes per burst period), and traced their path homologies at different stages of the memory spaces'
development. Once again, the dynamics of $\betaup_k(\mathscr{G})$ largely mirror those of $\betaup_k(
\mathfrak{G})$ and $\betaup_k(\mathcal{G})$, suggesting that the emerging episodic memory chains encoded by
place cell activity are structurally similar to cell assembly and neuronal chain order dynamics: disordered
at first, then consolidating over the learning period $T_{\min}^{(p)}(\mathscr{G})$.

It is worth noting that the pools of memory sequences could also be analyzed using singular-like homologies,
in terms of chains induced by mapping $\hat{\mathfrak{G}}_{\mathcal{T}}$ into $\mathscr{G}$, though we do
not pursue this approach here \cite{Grig3}.

\subsection{Topological consolidation}

As discussed in the Introduction, the hippocampus generates coarse representations of the external world---a
``subway map" of memorized experiences \cite{Parra,Balaguer}. Homological descriptions of this map are even 
coarser: the network's structure and spiking dynamics may change significantly without affecting the path or
simplicial homologies of the neuronal coactivity complex. For instance, alterations in the simplicial complex
$\mathcal{T}$, such as contractible protrusions, bulges, bends, or overall enlargements or shrinkages 
(Fig.~\ref{f3:env}C), remain indiscernible in homological analyses as long as no topological loops are 
created or eliminated. In other words, homological descriptions reflect only major, qualitative restructurings.
Yet physiological changes can be subtler, such as the consolidation of rapidly learned hippocampal memories
into longer-term, coarser-grained cortical structures that preserve the spatiality and overall morphology
of the original maps \cite{Audr,Wino1,Preston,Sekeres1,Robin,Sekeres2,Squire}.

The exact physiological mechanisms underlying these processes are unclear, but functionally, they involve
eliminating redundant or disused cell assemblies and consolidating inputs from active ones \cite{OReilly,Benna}.
Phenomenologically, such effects can be modeled by removing structurally redundant simplexes from the coactivity
complex $\mathcal{T}$, reducing the granularity of the corresponding finitary topological spaces while preserving
their homologies \cite{Grig5}. In particular, the maximally coarsened memory space---the \textit{core}, 
$\mathcal{C}(\mathcal{M}$)---captures the original topological structure using the smallest number of points and
neighborhoods (Fig.~\ref{f6:stng}A) \cite{Stng,Osk,McCord,Raume,Fth}. Interestingly, similar compact
representations are also discussed in neuroscience as Morris' ``location schemas" \cite{Tse1,Tse2,Morris,Farzanfar}.
In \cite{SchemaM}, it was argued that such schemas can be constructed as the cores of memory spaces produced
by place cells in a given environment, for specific physiological parameters of neuronal activity.

From the path-homological perspective, coarsening a memory space implies a reduction in the associated pool
of memory sequences, which could potentially disrupt the overall structure of serial activity. However, direct
computations reveal that the path homologies of the ``consolidated" set of sequences remain unchanged---that is,
a smaller population of heterologous sequences in reduced memory spaces preserve the structure of the original 
pool (Fig.~\ref{f6:stng}B). In other words, patterns of consolidated activity retain both their ordinal and spatial
organization. Biologically, this suggests that serial and spatial frameworks can be consolidated consistently.

Mathematically, the preservation of memory sequences' homologies results from the fact that spatial reduction
amounts to a homotopical transformation of the path complex. By this reasoning, one can use generic 
graph-homotopy retractions to model the consolidation of the sequential activity framework and the associated
decimations of the neuronal population. These transformations, by preserving path homologies, yield a smaller
repertoire of spiking activity that retains the original memory structure over smaller memory templates.

For instance, isolated sequences can be contracted to a single node, indicating that, at the level of 
abstraction used in this approach, solitary series of cognitive events may consolidate into a single episode 
(Fig.\ref{fig:hmt}A,B). Similarly, linear 'tendrils' extending from the graph can be retracted to its base, 
suggesting that sequences of extraneous cognitive episodes can be integrated into a compact core 
(Fig.\ref{fig:hmt}D). By this reasoning, sequential memory schemas can in general be modeled as retracts of 
the corresponding coactivity graphs.

\section{Discussion}
\label{sec:disc}

Representing spiking patterns by simplicial coactivity complexes achieves several objectives. First, it provides
a straightforward qualitative model of the cell assembly network---a generic schema that can incorporate numerous
cellular-level characteristics and reveal their system-level effects \cite{Eff,CAs,Rev,SchemaS}. Second, a 
coactivity $\mathcal{T}$ can be viewed as the \v{C}ech complex associated with the place field map. In morphing
environments, where the place field map deforms ``elastically" \cite{Gthrd,Ltgb,Wills,Trzky,Place,Blmnd,eLife},
its \v{C}ech complex remains intact, underscoring the topological aspect of the construction. Third, the 
simplicial semantics suggests structuring the network activity by identifying equivalent patterns of igniting
cell assemblies, interpreting the latter as homologous chains of simplexes. Given the hippocampus' role in the
topological aspect of spatial awareness, such computations may actually occur in the hippocampo-cortical network
\cite{Alv1,Poucet,Goodrich,Chen,Wu,Alv2,Alv3}.

On the other hand, recent studies portray the hippocampus as a universal generator of firing sequences, which
act as functional units for triggering cognitive episodes and behaviors \cite{Sosa,Deng,SqGen}. Experimental 
investigations into sequential neuronal activity reveal familiar patterns: the same firing order can adapt to 
different spatial and temporal scales, depending on specific tasks, contexts, or environments \cite{Terada,
	Maurer,Mau,Ezzyat}. Consequently, a particular configuration of firing sequences may correspond to a 
quasi-stable state of the underlying network, and represented by a path complex. The structure of this complex
is then derived by an alternative concept of equivalence among its components---path homology. 

This approach allows identifying the intrinsic structure of various sequence pools: igniting assemblies, firing
cells, and evoked memory episodes---regardless of the underlying physiological mechanisms. The findings suggest,
\textit{e.g.}, that hippocampal sequences generated by hundreds of cells and cell assemblies fall into only a 
few equivalence classes that collectively define the ordinal organization of hippocampal activity. Notably, this
organization remains stable across various spatiotemporal scales and is robust to geometric deformations of the
environment. It aligns with the retracting coactivity complex, pointing to the possibility of simultaneous 
consolidation of ordinal and spatial maps \cite{McKz1}. Path-homological analyses further indicate that ordinal
structures are predominantly supported by shorter sequences, much like spatial maps rely on the coactivity of
smaller cell groups \cite{CAs,Reimn}, suggesting that biological information is generally conveyed by small 
computational units. Finally, the ordinal organization of neuronal activity emerges and stabilizes over 
timescales comparable to those of spatial map learning, exhibiting similar dynamics.

Certain parallels between the simplicial and path-homological approaches are natural, as they ultimately provide
complementary descriptions of a unified information flow, share elements, and are functionally interconnected.
Indeed, hippocampal maps generally influence the spiking order (much like subway maps define the order in which
stations are visited), and, conversely, growing pools of interwoven sequences create nexuses of consistently
linked elements that yield connectivity maps (just as traversing enough stations reveals the overall metro plan).
Thus, it may appear that the differences between these approaches are merely semantic: if learning is
\textit{viewed} as spatial mapping, homologous spiking patterns would be identified ``simplicially," whereas if
the focus is on sequential experiences, like memorizing routes between locations, tracing path homologies among
sequenced activities is more appropriate. However, this is not the case---the two principles by which homologous
patterns are selected differ fundamentally, leading to qualitatively distinct frameworks: ordinal structures do
not generally reflect spatial affinities, and vice versa, and the timescales for establishing these structures
also differ. In other words, spatial and ordinal cognitive frameworks are not interchangeable: ``mapping out" a
new environment is distinct from memorizing routes between locations or internalizing other sequential experiences.
Thus, the ongoing ``space vs. sequences" debate in the hippocampal literature has a profound mathematical 
substance \cite{Vaz,SqGen,Dragoi3,Raju,Rueck}.

There are many examples of natural scientific systems, deriving their structure and dynamics from homological
and cohomological equivalences among their constituents \cite{Rabson,Stasheff}. Similar principle may apply to
neuronal systems, where the organization of spiking patterns based on simplicial or path-homological equivalences
can be imperative for large-scale, system-level behavior and be manifested at cognitive stage. At this point, it
remains \textit{a priori} unclear which homological framework best captures the essence of actual physiological
computations---this question must be addressed experimentally. Evidence of nearly simultaneously igniting cell
groups \cite{Wirth,Kolib,Cao,Kutter} and intrinsic serial firing motifs \cite{Yiu,Farooq} suggests that both 
principles could be implemented within the hippocampus or in supporting networks. In either case, the outlined
expansion of qualitative information-processing principles can be seen as a principal generalization of the 
hypothesized \textit{topological nature of hippocampal activity} proposed in \cite{eLife,PLoS,Rueck,Poucet}.

A practical implication of this conjecture is that the well-documented behavioral and cognitive effects produced
by topological restructuring of the environment \cite{Alv1,Poucet,Goodrich,Chen,Wu,Alv2,Alv3}, should also arise
in response to ordinal reorganization. In other words, qualitative changes in the serial organization of the 
environment, or externally induced alterations in neuronal spiking order, are anticipated to trigger consistent
and statistically significant comportmental responses, driven by the necessity to adapt to the change.

A practical implication of this conjecture is that the well-documented behavioral and cognitive effects induced
by topological restructuring of the environment \cite{Alv1,Poucet,Goodrich,Chen,Wu,Alv2,Alv3} should also emerge
in response to ordinal reorganization. In other words, qualitative changes in the sequential organization of the
environment or externally induced alterations in neuronal spiking order should be expected to elicit consistent
and statistically significant perceptual and adaptive responses, driven by the necessity to accommodate these 
changes.

\section{Acknowledgments}
\label{sec:ack}
We are grateful to Prof. A. Grogor'yan for the software used to compute the path homologies. Additional 
\textsf{MATLAB} software by Dr. M. Yutin is available
\hyperlink{https://github.com/SteveHuntsmanBAESystems/PerformantPathHomology}{SteveHuntsmanBAESystems}, 
see \cite{Chow}.

The work was supported by NSF grant 1901338 and partly by NIH grants R01AG074226 and R01NS110806.

\newpage
\section{Mathematical supplement: basic notions and glossary}
\label{sec:met}
\setcounter{equation}{0}

\textbf{1. Geometric simplexes and complexes}
\begin{itemize}[nosep,align=parleft,labelsep=-0.6em]
	\vspace{7pt}

	\item \textit{Geometric $d$-dimensional simplex} $\kappa^{d}$ is the convex hull of	its $(d+1)$ vertices
	(Fig.~\ref{f1:simp}A).
	\item The \textit{facet} opposite to a given vertex in a $d$-dimensional simplex $\kappa^{d}$ is itself a
	$(d-1)$-dimensional simplex $\kappa^{d-1}$ and vice versa, $\kappa^{d}$ can be produced by joining the 
	points of $\kappa^{d-1}$ to a new, $(d+1)^{\textrm{st}}$ vertex (Fig.~\ref{f1:simp}B). 
	The order of vertexes defines the \textit{orientation} of a simplex.
	\item \textit{Boundary of a $d$-simplex}, $\partial\kappa^{d}$, consists of its $(d-1)$-dimensional faces
	(Fig.~\ref{f1:simp}B).
	\begin{equation*}
		\partial\kappa^{d}=\kappa^{d-1}_1\cup\kappa^{d-1}_2\cup\ldots\cup\kappa^{d-1}_{d+1}.
		%\label{sbound}
	\end{equation*}
	\item A simplex is \textit{maximal}, if it is not a subsimplex of any other simplex.
	\item\label{gsimplx}
	\textit{Geometric simplicial complex} $K$ is a combination of properly assembled geometric simplexes that
	fit	each other vertex-to-vertex, side-to-side, with no overruns or other mismatches (Fig.~\ref{f1:simp}C).
	Formally, it is required that a non-empty intersection of any two simplexes $\kappa_1,\kappa_2$ in $K$ is 
	another simplex from $K$.
	\item\label{sk} The collection of all simplexes of dimensionality $d$ and less forms the \textit{$d$-skeleton}
	of a simplicial complex $\Sigma$, $sk_d(K)$.
\end{itemize}

\vspace*{7pt}
\textbf{2. Abstract simplexes and complexes}
\vspace{7pt}
	\begin{itemize}[nosep,align=parleft,labelsep=-0.6em]
	\item \label{asc} 
	Since all that matters for quantifying topology of simplicial complexes is how their simplexes match, it 
	is natural to ignore their ``filling" altogether and pass on to the notion of \textit{abstract
		simplexes}---ordered lists of simplexes' vertexes (\ref{simplex}), and their collections---\textit{
	abstract simplicial complexes}, $\Sigma$. An abstract simplex may be intuitively apprehended as the set
	of vertexes of a geometric simplex. 
	The ``face-matching" of	the geometric simplexes reduces to the requirement that if an abstract simplex 
	belongs to a simplicial	complex, then so do all its faces \cite{Alexandrov}.
	\end{itemize}

	\vspace*{7pt}
	\textbf{3. Graphs}
	\begin{itemize}[nosep,align=parleft,labelsep=-0.6em]
		\vspace*{7pt}
	\item \textit{Directed graph}, or \textit{digraph}, $G$ consists of vertices, $V=\{v_1,v_2,\ldots,v_n\}$,
	and directed edges between certain pairs of vertexes. Formally, $G$ is described by its connectivity matrix
		\begin{equation}
			C_{ij}(G)= \begin{cases}
				1, \,\,\, & \mbox{if $v_i$ is connected to $v_j$ by edge $e_{ij}$,} \\ 
				0 \,\,\, & \mbox{otherwise}.
			\end{cases}
			\nonumber
		\end{equation} 
	If the matrix $C_{ij}$ is symmetric, the graph is viewed as undirected, \textit{i.e.}, $e_{ji}$ always exists if
	$e_{ij}$ does.
	\item Any graph is a $1D$ simplicial complex.
	\item \textit{Digraph map} $f: G\to H$ sends vertexes $v_i$ of the digraph $G$ into the vertexes $w_p$ of
	the digraph $H$, so that each $G$-edge either maps on a $H$-edge, or squeezes into a $H$-vertex. 
	Incidentally, digraphs and maps between them form a category.
	\item $G$ is \textit{transitive}, if whenever edges $e_{ij}$ and $e_{jk}$ exist, then $e_{ik}$ is also an
	edge.
	\item \textit{Line digraph}, $I_n$, consists of $n$ ordered directed links between consecutive pairs of
	vertexes, e.g., $I_n=\{v_1\to v_2\leftarrow v_3\to\ldots\leftarrow\ v_{n-1}\leftarrow v_n\}$. 
	If the last vertex, $v_n$, connects to the first, $v_1$, then digraph is \textit{cyclic}, e.g., $S_n=\{v_n\to
	v_1\leftarrow v_2\to\ldots\leftarrow v_{n-1}\to v_n\}$.
	\item Linear digraphs of length one are the two \textit{directed dyads}: $I^{+}=(\varv\to\varv')$ and
	$I^{-}=(\varv\leftarrow\varv')$.
	\item \textit{Cartesian product} of a digraph $G$ with a vertex set $V$ and a digraph $H$ with its vertex
	set $W$, is a digraph $G\times H$ whose vertices have a $G$- and a $H$-component, $(v_i,w_p)$. Vertexes
	$(v_i,w_p)$ and $(v_j,w_q)$ are connected by an edge if either $v_i$ connects to $v_j$ in $G$, with the
	$H$-component fixed, $w_p=w_q$, or vice versa, $w_p$ connects to $w_q$ in $H$, with fixed $v_i=v_j$.
	\item \textit{Cylinder} over a digraph $G$ is the direct product, $G\times I_n$.
	\item \textit{Clique} of order $d$, $\varsigma^{(d)}$ is a fully interconnected subgraph with $(d+1)$
	vertexes $v_{i_0},v_{i_1},\ldots,v_{i_d}$. 
	\item \textit{Clique complex} associated with a graph $G$, $\Sigma_{G}$, is the full collection of $G$-cliques
	\cite{KnillE}. 
	\item A finite simplicial complex $\Sigma$ can be represented by the following constructions:
\vspace{7pt} 
\begin{enumerate}[i.,nosep,labelsep=0.6em]
	\item vertices of the \textit{simplex connectivity graph}, $G_{\Sigma}$, $v_{\sigma}$, represent simplexes
	of $\Sigma$, connected by an undirected edge $e_{\sigma\sigma'}$ if $\sigma$ and $\sigma'$ intersect, 
	$\sigma\cap\sigma'\neq\varnothing$ \cite{KnillE}. 
	\item vertexes $v_{\sigma}$ and $v_{\sigma'}$ of the \textit{maximal simplex connectivity graph}, $\mathfrak{G}
	_{\Sigma}$, are connected by an undirected edge $e_{\sigma\sigma'}$ if the maximal simplexes $\sigma$, 
	$\sigma'\in\Sigma$ overlap, $\sigma'\cap\sigma\neq\varnothing$.
	\item vertices $v_{\sigma}$ and $v_{\sigma'}$ of the \textit{simplex inclusion digraph}, $\mathsf{G}_{\Sigma}$,
	are connected by a directed edge $e_{\sigma\sigma'}$ if $\sigma'$ is a facet of $\sigma$, \textit{i.e.}, if $\sigma
	\subsetneq\sigma'$.
	\item If $\Sigma$ is a clique complex, then the \textit{barycentric refinement} of its $1D$ skeleton, $sk_1
	(\Sigma)=G$, is the graph whose vertexes correspond to cliques of $\varsigma$ of $\Sigma$, and the edge 
	$e_{\varsigma\varsigma'}$ between $\varsigma$ and $\varsigma'$ exists if $\varsigma\subsetneq\varsigma$. 
	Thus, barycentric refinement of $G$	is the connectivity graph of its clique complex $\Sigma_{G}$.
	\item\label{Ssig} vertexes $v_{\sigma}$ and $v_{\sigma'}$ of the \textit{face connectivity digraph}, 
	$\mathtt{G}_{\Sigma}$, connect by $e_{\sigma\sigma'}$ if $\sigma$ is a maximal facet of $\sigma'$, 
	$\dim(\sigma)=\dim(\sigma')+1$. 
	\item vertexes $v_{x}$ and $v_{x'}$ of the \textit{topogenous graph} $\mathscr{G}_{\Sigma}$ represent points
	in a topological space $X$, and the edges $e_{xx'}$ represent their adjacency.
\end{enumerate}
\end{itemize}
\vspace{7pt}

\textbf{4. A few bullet points of the simplicial homology theory}---for more details see \cite{Alexandrov,EdelHarer,
	ZmBook}.
\vspace{7pt}
\begin{itemize}[nosep,align=parleft,labelsep=-0.6em]
	\item \textit{Chains} of dimensionality $k$ are formal linear combinations of oriented $k$-dimensional 
simplexes, in which $\sigma^k_1$ is counted $m_1$ times, $\sigma^k_2$ is counted $m_2$ times, etc.,
\begin{equation*}
	c^k=m_1\sigma^k_1+m_2\sigma^k_2+...+m_q\sigma^k_q,
	%\label{chain}
\end{equation*}
with the coefficients from a field $\mathbb{K}$. If the coefficients come from the Boolean field $\mathbb{Z}_2$,
then the chain is simply a list of ``present" ($m_i=1$) and ``absent" ($m_i=0$) simplexes. The
chains can be added or subtracted from one another, as well as multiplied and divided by coefficients, thus
producing linear spaces, $C_k(\Sigma)$. Intuitively, each chain space is an algebraization of its $k$-dimensional
part, e.g., the dimensionality of $C_k$ is given by the number of the $k$-simplexes in $\Sigma$, 
\begin{equation*}
	\dim C_k=\{\#\sigma\in\Sigma:\dim(\sigma)=k\}.
\end{equation*}
	\item \textit{Boundary of a $k$-simplex}, algebro-topologically, is a $(k-1)$-dimensional \textit{chain of
		faces}, as defined by the formula (\ref{dsmplx}).
	For example, the boundary of an interval $\sigma_{12}$ oriented from vertex $\sigma_1$ to vertex $\sigma_2$
	is a formal alternated sum $\partial\sigma_{12}=\sigma_1-\sigma_2$. The boundary of a filled $2D$ triangle
	is an alternated sum of its three sides, $\partial\sigma_{012}=\sigma_{12}-\sigma_{02}+\sigma_{12}$, etc.
	\item \textit{Boundary of a chain} is a combination of the contributing simplexes' algebraic boundaries,
	\begin{equation*} 
		\partial c^k=b^{k-1}=m_1\partial\sigma^k_1+m_2\partial\sigma^k_2+...+m_q\partial\sigma^k_q,
		%\label{bound}
	\end{equation*}
	which is hence an element of the $C_{k-1}(\Sigma)$-space. The full collection of boundary chains $b^{k-1}$
	forms a subspace of $C_{k-1}(\Sigma)$, denoted by $B_{k-1}(\Sigma)$.
	\item \textit{Nilpotence}. By direct verification, $\partial^2c^k=0$, \textit{i.e.}, boundary of a boundary always
	vanishes.
	\item \textit{Cycles}. Not all boundaryless chains, $\partial c^k=0$, are boundaries---in general, they are
	combinations of simplex chains that ``loop onto themselves" in the complex. Such chains are referred to as
	\textit{cycles}, and denoted by $z$. Linear combinations of cycles of dimensionality $k$ also form a vector
	space, denoted by $Z_k(\Sigma)$, which is smaller than $C_{k}(\Sigma)$, but generally larger than $B_{k}(
	\Sigma)$ \cite{Alexandrov,EdelHarer,ZmBook}.
	\item \textit{Chain complex} is a system of spaces $C_k$, in which higher-order spaces are mapped into
	lower-order ones,
	\begin{equation*}
		0%\leftarrow \mathbb{K}
		\xleftarrow{\partial} C_0\xleftarrow{\partial}\ldots \xleftarrow{\partial} C_{k-1}\xleftarrow{\partial}
		C_k\xleftarrow{\partial}\ldots,
	\end{equation*}
	as facets of simplexes from $C_k$ embed into $C_{k-1}$. In other words, at each step the boundary operator
	$\partial$ maps $C_k$ into $B_{k-1}$ that lays within $C_{k-1}$.
	\item \textit{Homologous chains}. Adding or removing a $k$-dimensional boundary amounts to deforming a
	cycle $z$ by ``snapping" it over a series of $(k+1)$-dimensional simplexes and producing an equivalent, or
	\textit{homologous} cycle $z\pm\partial c=z'$ (Fig.~\ref{f1:simp}D). Cycles that cannot be matched by 
	adding or removing full boundaries are nonequivalent. Identifying equivalence classes hence amounts to
	factoring out the ``boundary parts," \textit{i.e.}, taking the quotient of the cycle space by the boundary space,
	$H_k(\Sigma) = Z_k(\Sigma)/B_k(\Sigma)$. The basis elements of the resulting space of $k^{\textrm{th}}$
	\textit{simplicial homology groups}, $H_k(\Sigma)$, represent $k$-dimensional loops, counted up to a 
	continuous deformation. 
	\item \textit{Betti numbers} are the dimensionalities of homology spaces, $b_k(\Sigma)=\dim H_k(\Sigma)$;
	these numbers count the topologically distinct $k$-loops in the complex $\Sigma$. For example, deforming a
	$0D$ chain (a combination of $0D$ points) amounts to ``sliding" those points inside of $\Sigma$; the 
	dimensionality of the corresponding homology space ($0^{\textrm{th}}$ Betti number), $\beta_0(\Sigma)$, is
	equal to the number of ``sliding domains" in $\Sigma$, \textit{i.e.}, the number of its connected components. If a
	simplicial complex comes in one piece, its $\beta_0(\Sigma)=1$. $\beta_1(\Sigma)$ equals to the number of
	holes in $\Sigma$. An example of a $2D$ noncontractible loop captured by $\beta_2$ is a ``hollow" 
	tetrahedron. Being a topological $2D$ sphere, it can hold no $1D$ loops and hence its Betti numbers are
	$\beta_0=1$, $\beta_1=0$, $\beta_2=1$, $\beta_{n>2}=0$. 
\end{itemize}
\vspace{7pt}
\textbf{5. Path homology theory} is outlined below following the publications \cite{Grig1,Grig2,Grig3,
	Grig4,Grig5}.
\vspace{7pt}
\begin{itemize}[nosep,align=parleft,labelsep=-0.6em]
	\item Let $V=\{v_1,\ldots, v_n\}$ a finite set. An \textit{elementary path} of length $k$ is an	arbitrary
	sequence of elements, $e_{i_0i_1\ldots i_k}$. These paths are basic units of the theory, analogous to
	simplexes of simplicial topology. 
	\item \textit{Path complex} $\mathcal{P}$ is collection of the elementary paths. Lengths of paths define
	their order, \textit{i.e.}, correspond to the dimensionality of simplexes. We consider only finite path 
	complexes that include paths up to some maximal length, $n$. If $P_l$ is the set of all elementary paths of
	length $l$, then the collection, $\mathcal{P}_k=P_0\cup P_1\cup P_2\cup\ldots\cup P_k$, is the $k$-skeleton
	of $\mathcal{P}$---the analogue of the $k$-skeleton $\Sigma_k$ of a simplicial	complex. 
	\item The paths forming a complex $\mathcal{P}$ must allow ``plucking" of the endpoints, \textit{i.e.}, if
	$e_{i_0i_1\ldots i_k}$ is in $\mathcal{P}$, then $e_{i_1\ldots i_k}$ and $e_{i_0i_1\ldots i_{k-1}}$ must also
	belong to $\mathcal{P}$ (Fig.~\ref{f2:de}B).
\end{itemize}
\vspace*{7pt}
\textbf{6. Algebraization of path chain complexes}
\vspace*{7pt}
\begin{itemize}[nosep,align=parleft,labelsep=-0.6em]
	\item \textit{Path chains} of order $k$ are formal linear combinations of elementary paths of length $k$,
	with the coefficients in a field $\mathbb{K}$, 
	\begin{equation*}
		\lambda=\sum_{i_0\ldots i_k\in V}\lambda^{i_0\ldots i_k}e_{i_0\ldots i_k}.
	\end{equation*}
	The collection of \textit{all} path $k$-chains forms a linear space $\Lambda_k$, e.g., $\Lambda_0$ are
	linear combinations of vertexes $e_i$, $\Lambda_1$, are the combinations of all pairs of vertices, $e_{ij}$,
	etc.
	\item \textit{Boundary operator} (\ref{dpath}) brakes the elementary paths into segments that precede the 
	$q^\textrm{th}$ step and the segments that follow it, e.g., $\partial e_{i}=e$, $\partial e_{ij}=e_j-e_i$,
	$\partial e_{ijk}=e_{jk}-e_{ik}+e_{ij}$, etc. By linearity, the boundary of a generic path chain is
	\begin{equation*}
		\partial \lambda=\sum_{i_0,\ldots,i_k}\lambda^{i_0\ldots i_k}\partial e_{i_0\ldots i_k}=\sum_{i_0
			\ldots i_k}\sum\limits_{q=0}^{k}(-1)^{q}\lambda^{i_0\ldots i_k} e_{i_0\ldots\cancel{i}_q\ldots i_k}.
	\end{equation*}
	\item By direct verification, $\partial^2\lambda=0$ for any $\lambda$, which implies that the boundary 
	operator (\ref{dpath}) induces a chain complex $\Lambda_{\ast}(\mathcal{P})$,
	\begin{equation*}
		0%\xleftarrow{\partial} \mathbb{K} 
		\xleftarrow{\partial} \Lambda_0\xleftarrow{\partial}\ldots \xleftarrow{\partial}\Lambda_{k-1}
		\xleftarrow{\partial} \Lambda_k\xleftarrow{\partial}\ldots \, ,
	\end{equation*}
	in which the path-chains nullified by $\partial$ are the path-cycles, $\partial\zeta=0$ and the chains of
	the form $\beta=\partial\lambda$ are the boundaries. As with the simplicial complexes, both types of chains
	form their respective linear spaces, $Z_k(\mathcal{P})$ and $B_k(\mathcal{P})$. The factor space of the 
	path-cycles over the path-boundaries yield the \textit{path homology space}, $\HP_{k}(\mathcal{P})$, 
	analogous to the simplicial homologies of simplicial complexes $H_{k}(\Sigma)$.
	\item If a path complex is formed by a collection of certain select, or \textit{allowed} elementary paths,
	then their linear combinations,
	\begin{equation*}
		\Alpha_k(\mathcal{P})=\left\{\sum_{j_0\ldots j_k}\alpha^{j_0\ldots j_k}e_{j_0\ldots j_k}\right\},
	\end{equation*}
	form a subspace of $\Lambda_k$. Examples:
	\vspace{7pt} 
		\begin{enumerate}[i.,nosep,labelsep=0.6em]
		\item In a graph-representable complex, allowed paths are the ones that run along the graph's edges,
		other sequence are excluded.
		\item The allowed paths on the (maximal) simplex connectivity graph are the sequences of overlapping
		(maximal) simplexes.
		\item From a biological perspective, if $V$ is the full set of cells or assemblies in a given network,
		the allowed vertexes, $P_0$, may be the ones that exhibit activity in a given environment. The allowed
		links, $P_1$, may be the ones that represent synaptic connections, rather than all pairs of coactive
		cells, $P_2$, etc., may be synaptically interconnected triples of neurons, etc.
	\end{enumerate}
	\vspace{7pt}
	\item In order for the boundary operator (\ref{dpath}) to act on $A_k$, the boundaries of the allowed paths
	should also be allowed,
	\begin{equation*}
		\partial \Alpha_k\subset \Alpha_{k-1}.
		%\label{AA}
	\end{equation*}
	However, in contrast with the simplicial case, where each term in the right hand side of (\ref{dsmplx})
	is a part of the complex $\Sigma$ by design, terms appearing in the right hand sides of (\ref{dpath}) may
	structurally fall out of $\mathcal{P}$ (Fig.~\ref{f2:de}C). Yet, there exists a subclass of the allowed
	path chains---the \textit{operational path-chains} that	have allowed boundaries,
	\begin{equation*}
		\Omega_k=\{\alpha\in \Alpha_k:\partial \alpha\in\Alpha_{k-1}\}.
		%\label{Omdef}
	\end{equation*}	
	A simple, but fundamental property of such paths is that the boundary operator (\ref{dpath}) acts on them
	without fallacies. Indeed, if $\alpha\in\Omega_k$ then $\partial\alpha\in\Alpha_{k-1}$, and $\partial^2
	\alpha=0 \in\Alpha_{k-2}$, which means that $\partial\alpha\in \Omega_{k-1}$. Thus, $\Omega$-paths form
	their own chain complex, $\Omega_{\ast}$,
	\begin{equation*}
		0\xleftarrow{\partial}\mathbb{K}\xleftarrow{\partial}\Omega_0\xleftarrow{\partial}\ldots
		\xleftarrow{\partial}\Omega_{k-1}\xleftarrow{\partial}\Omega_k\xleftarrow{\partial} \Omega_{k+1}
		\xleftarrow{\partial} \ldots,
	\end{equation*}
	whose homology groups (the \textit{reduced path homology groups} of $\mathcal{P}$) define the structure
	of the allowed paths.
	\item Digraphs and digraph maps form a category, distinct from the category of simplicial complexes and
	simplicial maps.
	\item A digraph map $f: G\to H$ induces a homomorphism of operational chain complexes,
	\begin{equation*}
		f_{\ast}|_{\Omega_{p}(G)}: \Omega_{\ast}(G,\mathbb{K})\to \Omega_{\ast}(H,\mathbb{K}),
	\end{equation*}
	and of path homology groups,
	\begin{equation*}
		f_{\ast}: \HP_{\ast}(G,\mathbb{K})\to \HP_{\ast}(H,\mathbb{K}).
	\end{equation*}
	Thus, path homologies can be used to classify digraphs relative to such maps.
\end{itemize}
\vspace*{7pt}
\textbf{7. Simplicial vs. path homologies}
\vspace*{7pt}

\begin{itemize}[nosep,align=parleft,labelsep=-0.6em]
	\item A path complex $\mathcal{P}$ is:
	\vspace{7pt}
	\begin{enumerate}[i.,nosep,labelsep=0.6em]
		\item\textit{perfect}, if it contains all subpaths of its elementary paths, so that $\mathcal{P}_k
		=\Alpha_k(\mathcal{P})=\Omega_k(\mathcal{P})$. In such complexes, the boundary (\ref{dpath}) is 
		structurally identical to (\ref{dsmplx}).
		\item \textit{monotone}, if its vertices can be numbered such that the numbering increases along
		each elementary path in $\mathcal{P}$. In a digraph, this implies that all edges comply with the		
		ordering. For example, if a path complex contains two elementary paths with conflicting vertex order,
		e.g., $e_{i_0,i_3,i_5,i_2}$ and $e_{i_2,i_7,i_5}$ have a conflicting order of $i_2$ and $i_5$, then
		this complex clearly can not be monotone. Conversely, any finite complex without such conflicts is
		monotone.
		\item perfect and representable by a digraph if and only if the latter is \textit{transitive}.
	\end{enumerate}
	\vspace{7pt}
	\item $\mathcal{P}$ is the path complex of a simplicial complex if and only if it is perfect and monotone.
	This combination is rare in applications, e.g., the neuronal coactivity graph or the synaptic connectivity
	graph are typically not transitive. To build a positive example, consider elementary paths that pass through
	vertexes enumerated in increasing order. The subsequences of such paths will also be ordered, \textit{i.e.}, the 
	complex $\mathcal{P}_G$	will be perfect and representable by a transitive graph $G(\Sigma)$, and the 
	homologies $\HP_k(G)$ and $H_k(\Sigma)$, $k\geq0$, will be isomorphic \cite{Grig1,Grig2}.
\end{itemize}
\vspace{7pt}
\textbf{8. Regularization of path complexes}
\vspace{7pt}
\begin{itemize}[nosep,align=parleft,labelsep=-0.6em]	
\item\textit{Regular elementary paths} are the ones that contain no consecutively repeating indexes: in
$e_{i_0\ldots i_n}$, $i_{k-1}\neq i_k$, for all $k$. The chain spaces $\Lambda_k$ can be decomposed into the
regular and the irregular components, $\Lambda_k=R_k\bigoplus I_k$, which, however, do not produce separate
path complexes, because $R_k$ and $I_k$ are not, by themselves, invariant with respect to the boundary
operator (\ref{dpath}). For example, the boundary of a regular path $e_{iji}$,
\begin{equation*}
	\partial e_{iji}=e_{ji}-e_{ii}+e_{ij},
\end{equation*}
includes an irregular segment $e_{ii}$. It turns out however, that one can build a homological classification
of the regular paths simply by discounting ``irregularities," \textit{i.e.}, by viewing two path chains as equivalent,
if they differ by an irregular path. In other words, one can use
\begin{equation*}
	\tilde{\partial} e_{iji}=e_{ji}+e_{ij}
\end{equation*}
instead of the ``na\"ive" $\partial e_{iji}$ above. The resulting equivalence classes contain a unique regular
path each, \textit{i.e.}, any path can be reduced to its unique regular representative---\textit{regularized}.
This procedure also induces a regularized boundary operator $\tilde{\partial}$, which allows defining the 
\textit{regular chain complex}
\begin{equation*}
	0\xleftarrow{\tilde{\partial}} R_0\xleftarrow{\tilde{\partial}} \ldots \xleftarrow{\tilde{\partial}}
	R_{k-1}\xleftarrow{\tilde{\partial}} R_k\xleftarrow{\tilde{\partial}} \ldots \ \ .
	%\label{cR}
\end{equation*}
whose homologies classify regular paths \cite{Grig1}. 
\item The set of allowed regular $k$-path is a subset of the total set of regular $n$-paths. The boundary
operator $\tilde{\partial}$ acting on such paths defines regular operational chains,
\begin{equation*}
	\tilde{\Omega}_k=\{v\in \Alpha_k:\partial v\in A_{k-1}\},
\end{equation*}
which are characterized by their regular homologies. 
\item A path complex $\mathcal{P}$ is \textit{strictly regular} if it is regular and does not breed irregularity,
\textit{i.e.}, contains no paths of the form $e_{...iji...}$. For example, a path complex of a simplicial complex is
strictly regular because its paths' indices are strictly increasing. The path complex of a digraph is strictly
regular iff there are no loops (no $e_{ii}$ edges and no simultaneously present $e_{ij}$ and $e_{ji}$ edges).
The corresponding homologies are the only ones used in our analyses.
\end{itemize}

\vspace{7pt}
\textbf{9. Homotopies}
\vspace{7pt}
\begin{itemize}[nosep,align=parleft,labelsep=-0.6em]
	\item Two digraph maps $f,g: G\to H$ are \textit{homotopic}, $f\simeq g$, if there is a map $F: G\times
	I_n\to H$ such that
	\begin{equation*}
		F|_{G\times \{v_0\}}=f\ \ \text{and}\ \ F|_{G\times \{v_n\}}=g.
		%\label{Ffg}
	\end{equation*}
	In particular, $f$ and $g$ are one-step homotopic ($n=1$), $f\eqcirc g$, if $f(v_i)$ and
	$g(v_i)$ either connect by an edge or coincide for all $v_i$.
	\item Any homotopy, $f\simeq g$, amounts to a finite sequence of one-step homotopies, $(f=f_0)\circ f_1
	\circ\ldots\circ (f_n=g)$, where $f_0=id$ is the identity map, and $f_k\eqcirc f_{k+1}$.
	\item Two digraphs $G$ and $H$ are \textit{homotopy equivalent}, $H\simeq G$, if there exist digraph maps
	$f:G\to H$ and $g:H\to G$, such that
	\begin{equation*}
		f\circ g\simeq id_H,\ \ \ \ \ g\circ f\simeq id_G.
	\end{equation*}
	The maps $f$ and $g$ are then \textit{homotopy inverses} of each other.
	\item Homotopy equivalent digraphs have isomorphic path homologies.
	Homotopic maps induce the \textit{identical} homomorphisms of path homologies.
	\vspace{7pt}
	\begin{enumerate}[i.,nosep,labelsep=0.6em]
		\item A cylinder over a digraph $G$ is homotopic to its base, $G\times I_n\simeq G$, for any $I_{n\geq
			 0}$.
		\item Natural inclusion of $G$ into a cylinder $G\times I_n$ over it, $i_{k}: v\to (v,\iota_k)$, and the 
		projection from a cylinder $G\times I_n$ to its base, $p:(v,\iota_k)\to v$, induce isomorphism of path 
		homologies.
	\end{enumerate}
	\vspace{7pt}
	%%%%%%%%%%%%%%%%%%%%%%%%%%%%%%%%%%%%%%%%
	
	\begin{figure}[h]
		%\begin{wrapfigure}{c}{0.65\textwidth}
		\centering 
		\includegraphics[scale=0.84]{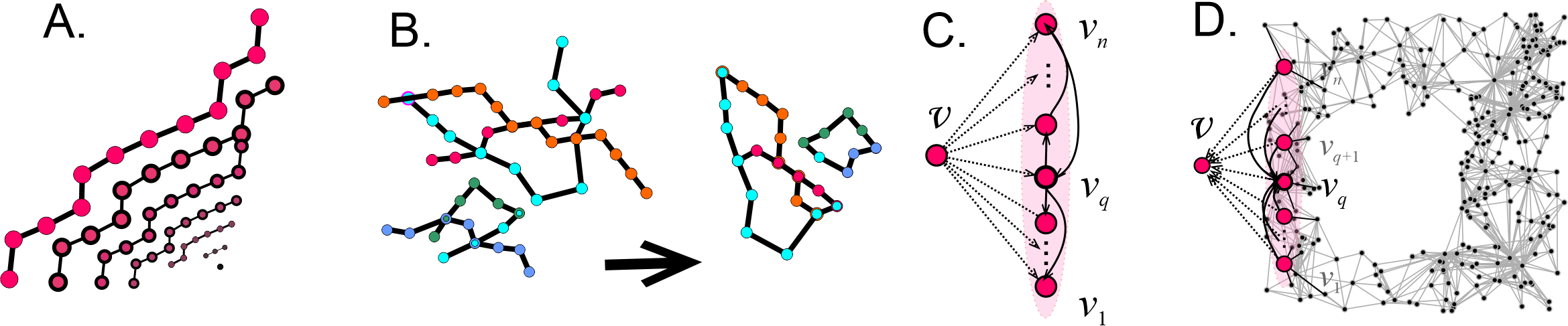}
		\caption{{\footnotesize
				\textbf{Graph homotopies}.
				\textbf{A}. Linear path can be contracted into a single node. 
				\textbf{B}. ``Tendrils" extending out of generic path complexes can be retracted.
				\textbf{C}. Star graph: if a single ``apex" $\varv$ connects to all other vertexes, $v_1,v_2,
				\ldots,v_n$, either by outgoing (shown) or by incoming edges, then the entire graph can be
				contracted to $\varv$, \textit{i.e.}, its path homologies vanish. 
				\textbf{D}. If a vertex $\varv$ connects to a set of vertexes $v_1,v_2,\ldots,v_{k}$, and if one
				of the latter vertices, e.g., $v_{s}$, connects to the rest of them just as $\varv$ does (\textit{i.e.}, 
				$e_{v_s v_p}$ exists when $e_{\varv v_p}$ does and is similarly directed, $p=1,\ldots,k$) then 
				$\varv$ with all of its adjacent edges can be removed from the graph $G$ without altering its 
				homologies.
		}}
		\label{fig:hmt}	
		%\end{wrapfigure}
	\end{figure}
	
	%%%%%%%%%%%%%%%%%%%%%%%%%%%%%%%%%%%%%%%
	\item \textit{Retraction} of a digraph $G$ onto a subgraph $H$ is a map $f:G\to H$ that fixes all the 
	vertices of $H$. A \textit{deformation retraction} is the one that can be homotopically deformed to the
	identity map. This implies that $G$ and $H$ are homotopically equivalent. Any deformation retraction is
	a composition of one-step retractions. A one-step retraction is a digraph map that fixes some vertices and
	moves all others either one step along the edges, or one step against the edges simultaneously.	
	\item A digraph $G$ is \textit{contractible} if it is homotopic to a single vertex. All the homology groups
	of a contractible graph are trivial, except for $\HP_0(G)=\mathbb{K}$ \cite{Ivash}.
	\vspace{7pt}
	\begin{enumerate}[i.,nosep,labelsep=0.6em]
		\item A connected tree can be contracted to its root vertex by ``trimming" the outer vertices. The process
		of removing the terminal vertexes at each branch with its adjacent edge is a deformation retraction. 
		\item A \textit{star-like} digraph has a vertex $\varv$ that connects to all other vertexes by linear
		digraphs. Such a graph is contractible to $\varv$, e.g., a facet $\sigma^{(2)}$ of a tetrahedron
		$\sigma^{(3)}$ can be contracted to its opposite vertex. 
		\item A $k$-dimensional cube $I^{k}_n=I_n\times I_n\times\dots\times I_n$ (direct product of $k\geq 1$
		$I_n$-lines), can be retracted to each of its facets, $I^{k}_n\simeq I^{k-1}_n$, and ultimately 
		contracted.
		\item Cycling triangle and a cycling square are contractible. Longer cycles, $S_{n>4}$, loose
		contractibility: $H_1(S_{n>4},\mathbb{K})\cong \mathbb{K}$. Furthermore, two unequal cycles, $S_{n>4}$
		and $S_{m>4}$, $n\neq m$, are homotopy inequivalent. 
	\end{enumerate}
	\item If $r:G\to H$ is a deformation retraction, then $G$ and $H$ are homotopy equivalent and the map $r$
	is homotopy inverse of $i$ and vice versa. 
	\item If a vertex $\varv$ connects to a group of other vertexes, $v_1,\ldots,v_n$, among which there is a
	special vertex, $v_p$, that connects to all the remaining $v$s whenever $\varv$ does ($\varv\to v_i
	\implies v_p\to v_i$, then removing $\varv$ along with all of its adjacent edges is a homology-preserving
	deformation retraction. This procedure may be viewed as removing this special kind of an ``embedded star"
	from $G$. The arrows in this construction can be simultaneously reversed ($\varv\leftarrow v_i\implies v_p
	\leftarrow v_i$). 
\end{itemize}
	
\vspace{7pt}
\textbf{10. The topological complexity} of a path-connected topological space $X$, $\textsf{TC}(X)$, is defined
as the smallest number of open sets, $U_1,U_2,\ldots U_k$, required to cover $X\times X$, such that each set 
$U_i$ allows for continuous motion planning from a starting to an ending point within its domain \cite{Farber}.
The value $\textsf{TC}(X)$ is a homotopy invariant \cite{Schwarz}, and for the spaces discussed above $\textsf
{TC}(\mathcal{E}^{(2)}_1)=\textsf{TC}(\mathcal{E}^{(3)}_1)=\textsf{TC}(\mathcal{E}^{(4)}_1)=2$.

\end{document}